\documentclass[aps,superscriptaddress,twocolumn]{revtex4-2}

\usepackage[T1]{fontenc}

\usepackage{graphicx}
\usepackage{dcolumn}
\usepackage{array}
\usepackage{tabularx}

\usepackage{floatrow}

\usepackage{eucal}
\usepackage[dvips]{epsfig}
\usepackage{amssymb}
\usepackage{amsfonts}
\usepackage{multirow}
\usepackage{chngcntr}

\usepackage[parfill]{parskip}    

\usepackage{amsmath,amsthm}
\usepackage{cleveref}
\usepackage{siunitx}

\usepackage{comment}

\newcommand{\be}{\begin{eqnarray}}
\newcommand{\ee}{\end{eqnarray}}
\newcommand{\bse}{\begin{subequations}}
\newcommand{\ese}{\end{subequations}}

\newcommand{\bnum}{\begin{enumerate}}
\newcommand{\enum}{\end{enumerate}}

\newcommand{\bit}{\begin{itemize}}
\newcommand{\eit}{\end{itemize}}

\newcommand{\bc}{\begin{cases}}
\newcommand{\ec}{\end{cases}}

\newcommand{\bpm}{\begin{pmatrix}}
\newcommand{\epm}{\end{pmatrix}}

\newcommand{\bvm}{\begin{vmatrix}}
\newcommand{\evm}{\end{vmatrix}}

\newcommand{\f}{\frac}

\newcommand{\tn}{\textnormal}

\usepackage{xcolor}

\DeclareSIUnit\chains{chains}
\DeclareSIUnit\filaments{filaments}

\begin{document}

\title{
Fluid flow generates bacterial conjugation hotspots by increasing the rate of shear-driven cell-cell encounters
}

\author{Matti Zbinden}
\affiliation{Institute of Environmental Engineering, Department of Civil, Environmental and Geomatic Engineering, ETH Zurich, Zurich, Switzerland}

\author{Jana S. Huisman}
\affiliation{Department of Physics, Massachusetts Institute of Technology, Cambridge, Massachusetts, USA}

\author{Natasha Blitvic}
\affiliation{School of Mathematical Sciences, Queen Mary University of London, UK}

\author{Roman Stocker}
\affiliation{Institute of Environmental Engineering, Department of Civil, Environmental and Geomatic Engineering, ETH Zurich, Zurich, Switzerland}

\author{Jonasz S\l{}omka}
\email{Corresponding author: jslomka@ethz.ch}
\affiliation{Institute of Environmental Engineering, Department of Civil, Environmental and Geomatic Engineering, ETH Zurich, Zurich, Switzerland}

\date{\today}
\begin{abstract}
Conjugation accelerates bacterial evolution by enabling bacteria to acquire genes horizontally from their neighbors. Plasmid donors must physically encounter and connect with recipients to allow plasmid transfer, and different environments are characterized by vastly different encounter rates between cells, based on mechanisms ranging from simple diffusion to fluid flow. However, how the environment affects the conjugation rate by setting the encounter rate has been largely neglected, mostly because existing experimental setups do not allow for direct control over cell encounters. Here, we describe the results of conjugation experiments in \textit{E. coli} in which we systematically varied the magnitude of shear flow using a cone-and-plate rheometer to control the encounter rate. We discovered that the conjugation rate increases with shear until it peaks at an optimal shear rate ($\dot{\gamma}=\SI{1e2}{\per\s}$), reaching a conjugation rate five-fold higher than the baseline set by diffusion-driven encounters. This optimum marks the transition from a regime in which shear promotes conjugation by increasing the rate of cell-cell encounters to a regime in which shear disrupts conjugation. Regions of high fluid shear are widespread in aquatic systems, in the gut of host organisms, and in soil, and our results indicate that these regions could be hotspots of bacterial conjugation in the environment.
\end{abstract}

\maketitle
\textbf{Significance statement}

Bacterial conjugation is a process in which bacteria exchange DNA upon contact. Different environments present cells with different encounter rates, setting physical limits on the conjugation rate, but how encounters shape conjugation remained unknown. By precisely stirring bacterial suspensions in a rheometer, we controlled the cell-cell encounter rate and examined how it affects the conjugation rate. Our experiments revealed that optimal stirring increases the conjugation rate five-fold, demonstrating that fluid flow can generate hotspots of conjugation in many environmental settings. Using encounter models, we predict that the ocean surface layer could be a hotspot of conjugation between planktonic cells because the turbulence there is strong enough to increase cell encounters, but still weak enough not to impair conjugation.

\textbf{Main text}

\begin{figure*}[t!]
    \centering
    \includegraphics[width=1.0\textwidth]{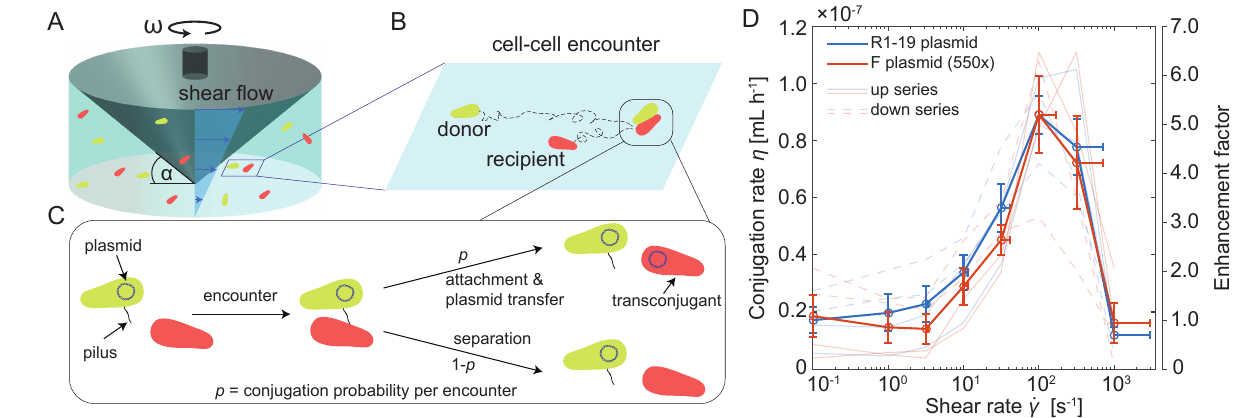}
    \caption{
\textbf{Bacterial conjugation rate peaks at an optimal shear rate.} 
    \textbf{A}, Schematic showing a suspension of plasmid donors and recipients in the cone-and-plate rheometer. The cone geometry generates a shear flow throughout the sample, which is controlled by the rotation rate $\omega$ of the cone. The angle $\alpha$ of the cone with the horizontal plate is exaggerated for visualization~(in reality $\alpha=4^{\circ}$).
    \textbf{B}, Cell-cell encounters in a fluid are generated by two mechanisms: (i) the shear flow, which induces relative movement between cells occupying adjacent vertical positions, and (ii) simple diffusion (Brownian motion). The experimental encounter rate is controlled by precisely controlling the applied shear rate.
    \textbf{C}, Schematic of the conjugation process, in which an encounter between donor and recipient leads to conjugation with probability $p$.
    \textbf{D}, The observed conjugation rate $\eta$ for \textit{E. coli} strains (TB204 for donors; NCM $\Delta$mot for recipients) sharing a conjugative plasmid (R1-19, blue; F, red) as a function of the applied shear rate $\dot\gamma$. Points represent the averages and vertical bars standard errors for $n=4$ independent up/down series (shaded lines) experiments for each plasmid (Methods). Each series represents a sequence of eight separate measurements with different aliquots at different shear rates (changing the shear rate from low to high in the up series and from high to low in the down series). The one-sided horizontal error bars represent an additional shear rate above the baseline shear rate~($\dot\gamma=\omega/\alpha$) generated by the secondary and turbulent flows at high rotation rates of the cone~(SI Section 1). Upon rescaling (550x) of the conjugation rate for the F plasmid, both plasmids show very similar shear responses, with the conjugation rate peaking at an optimal shear where the conjugation rate is approximately five-fold higher than the baseline. The y-axis on the right represents the enhancement factor computed by normalizing the conjugation rate for the two plasmids by the plateau value (the mean conjugation rate at the lowest shear for the R1-19 plasmid). 
    }
    \label{fig:Fig1}
\end{figure*}

Bacteria evolve rapidly because they share genes not only vertically, from parents to offspring, but also horizontally, between cells in the same environment~\cite{Aminov2011, Thomas2005,Soucy2015}. Up to 25\% of the bacterial genome consists of horizontally acquired genes~\cite{Nakamura2004}. Conjugation is a major pathway of horizontal gene transfer~\cite{Soucy2015}, takes place in a broad range of environments including the oceans~\cite{dahlberg1998situ,Petersen2019}, guts~\cite{Bakkeren2019} and soil~\cite{richaume1989influence}, and is strongly implicated in the global spread of antimicrobial resistance~\cite{Castaneda-Barba_NMR2024}. Conjugation starts with an encounter between a plasmid donor and a recipient cell, which leads to a mating junction established by a conjugative pilus. This is a type-IV secretion system (T4SS) that transports a single-stranded DNA copy of a plasmid from the donor to the recipient~\cite{arutyunov2013f} within approximately ten minutes~\cite{Babic_Science2008,Goldlust2023}. 
Decades of study have shown that the success of conjugation depends on many environmental and biological factors~\cite{LederbergTatum_Nature1946,sorensen2005studying, Sheppard2020,Benz_etal_ISME2020}. Observed conjugation rates vary by more than ten orders of magnitude and depend on plasmid type and size, pilus type, phylogenetic relatedness of donor and recipient, and the environment (temperature, medium, and whether cells are in liquid or on a surface)~\cite{Sheppard2020}.
Upon receiving the plasmid, the recipient becomes a transconjugant, capable of spreading the plasmid further at a rate limited by the encounter rate with other cells.

Different environments present plasmid donors with vastly different encounter rates with recipients. Environmental flows bring cells together through the stirring action resulting from fluid shear (i.e., gradients in fluid velocity), which creates relative movement between a donor and a recipient, or through diffusion (Brownian motion)~\cite{slomka2023encounter}. For example, due to turbulence, the surface ocean generates more cell-cell encounters than the quiescent deeper ocean~\cite{Franks2022}. Similarly, gut peristalsis~\cite{cremer2016effect} and preferential flow paths in porous media~\cite{kurz2022competition} generate high levels of intermittent stirring, with the associated shear rates spanning multiple orders of magnitude~\cite{kurz2022competition,schutt2022simulating}. Experiments have shown that moderate shaking of flasks can increase plasmid transfer, whereas vigorous shaking or vortexing decreases it~\cite{Zhong2010,Patkowski2023}. These results suggest that fluid motion can have a complex and, to date, poorly quantified effect on the conjugation rate. It has long been recognized that disentangling the encounter between donor and recipient from the plasmid transfer is required for a consistent comparison of the transfer efficiency of different plasmids or in different environments~\cite{sorensen2005studying}, but this distinction has proven very challenging to make when measuring conjugation in conventional shakers, because of the poor level of control over the flows that these create. Consequently, existing laboratory measurements cannot be used to systematically determine the effect of flow on conjugation.

Here, we quantitatively study the effect of fluid motion on the conjugation rate using a new assay in which the rate of cell-cell encounters is directly controlled.
Using a cone-and-plate rheometer, we systematically exposed a suspension of plasmid donors and recipients to a range of controlled flows with a defined shear rate, then counted transconjugants to study how the shear rate impacts the conjugation rate. We discovered that as the shear rate increases, the conjugation rate first stays constant, then increases up to a maximum, and then decreases. At the optimal shear rate, the conjugation rate is enhanced fivefold over the no-flow case in which Brownian motion alone drives conjugation. We use encounter rate theory and hydrodynamics to explain the existence of an optimal shear rate -- that is, an intermediate level of fluid motion for which conjugation is maximal -- by modeling the opposing effects of shear on conjugation. Our calculations show that, as the shear rate increases, the encounter rate between donors and recipients increases, but cells have increasingly less time to attach to each other and must then withstand increasingly higher shear forces that act to separate them. 

Our results indicate that flow in many natural environments, including aquatic systems, the gut of hosts, and soils, could locally lead to a major increase in conjugation rates. As an example, we predict that the ocean surface layer, particularly in breaking waves, can act as a hotspot of conjugation by generating shear rates high enough to enhance cell-cell encounters without impairing conjugation.

\subsection*{Results}

To quantify the effect of fluid shear on conjugation rates, we developed a new experimental assay that employs a cone-and-plate rheometer as a \lq cell collider\rq~(Fig.~\ref{fig:Fig1}). In this setup, a suspension of donor and recipient cells is placed between a plate and an inverted rotating cone~(Fig.~\ref{fig:Fig1}A). Rotation of the cone creates a shear flow in the suspension so that cells at different heights above the plate travel at different speeds. In this setup, two mechanisms bring cells together: shear flow and diffusion~(Fig.~\ref{fig:Fig1}B). Diffusion generates cell-cell encounters via Brownian motion of both donor and recipient cells. The shear flow generates encounters by creating a relative horizontal velocity $\delta v\sim \delta z \dot\gamma$ between cells separated vertically by a small distance $\delta z$. Here $\dot\gamma$ is the shear rate, which quantifies the velocity gradient. 
In the rheometer, the baseline shear rate $\dot\gamma$ is controlled by the angular speed $\omega$ of the rotating cone, so that $\dot\gamma=\omega/\alpha$, where $\alpha=4^{\circ}$ is the angle the cone makes with the horizontal plate~(Fig.~\ref{fig:Fig1}A). The cone base diameter is 4 cm, and the device holds a sample volume of 1.2 mL. Flows in such geometries have been studied in detail both numerically~\cite{fewell1977secondary} and experimentally~\cite{sdougos1984secondary}. As the angular speed increases, the flow is laminar for shear rates smaller than $\dot\gamma\approx\SI{35}{\per\second}$, then secondary flows develop, and the flow becomes turbulent for shear rates above $\dot\gamma\approx\SI{300}{\per\second}$~\cite{sdougos1984secondary}~(SI Section 1). Thus, at a low shear rate, the shear rate is constant throughout the sample and equal to the baseline shear rate. As the shear rate increases, the secondary and turbulent flows create regions where the shear rate is higher than the baseline value. 
Throughout the paper, we will use the baseline shear rate ($\dot\gamma=\omega/\alpha$) and use horizontal error bars to indicate additional shear present in the rheometer, estimated based on previous numerical simulations~\cite{fewell1977secondary} and torque measurements~\cite{sdougos1984secondary} (SI Section 1). The control of the shear flow achieved in the rheometer is a key feature of our approach, as it enables us to control the encounter rate between cells, in contrast to the complex flow occurring in conventional shakers. Upon encounter, cells may attach and conjugate or separate without transferring the plasmid~(Fig.~\ref{fig:Fig1}C).

\begin{figure*}[t!]
    \centering
    \includegraphics[width=1.0\textwidth]{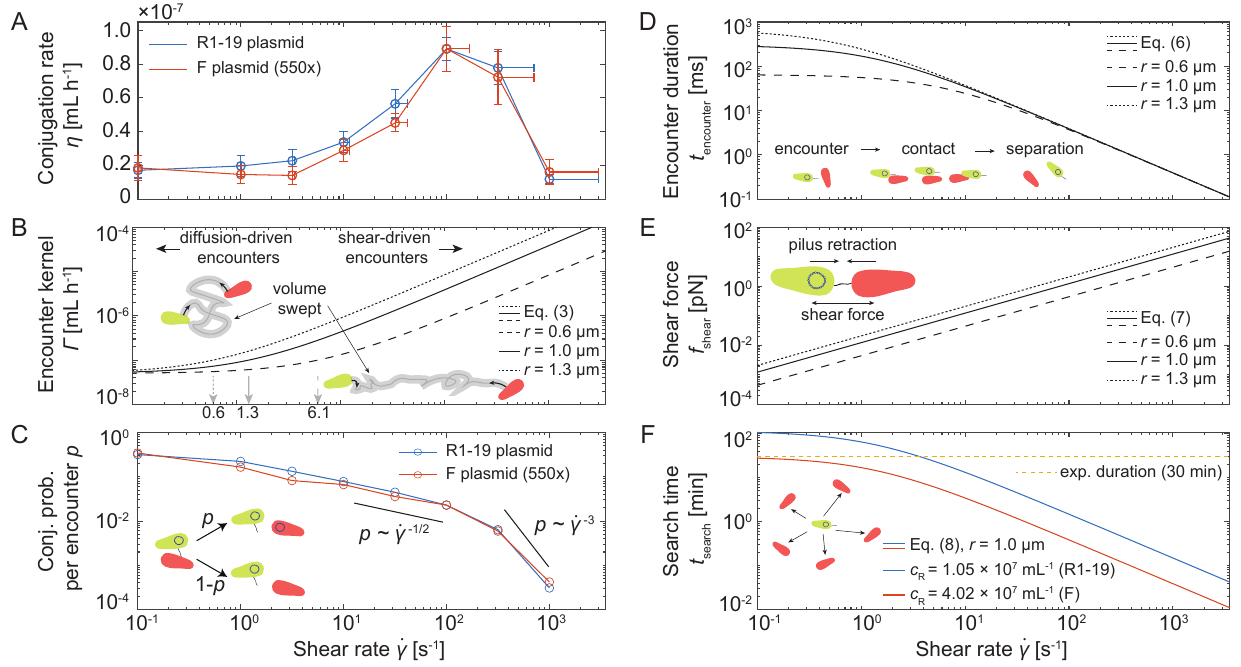}
    \caption{
\textbf{The conjugation rate depends on the physical mechanisms that bring cells together and on the shear forces that separate them.}
\textbf{A}, Observed conjugation rate between two \textit{E. coli} strains as a function of the applied shear rate, for two different conjugative plasmids (R1-19, blue; F, red; same data as in Fig. 1D).
\textbf{B}, Encounter kernel as a function of shear rate (Eq.~\ref{eq:gamma_tot}) for three different cell radii. Diffusion dominates encounters at low shear rates, whereas shear is the primary encounter mechanism at higher shear rates. The insets show schematics of trajectories in the two regimes, with the grey region indicating the volume swept in each case. The vertical arrows mark the shear rate values above which encounters are driven by shear for the different cell radii.
\textbf{C}, Probability of conjugation per cell-cell encounter as a function of shear rate (Eq.~\ref{eq:eta_p_Gamma}). This probability is obtained by dividing the observed mean conjugation rate (panel A) by the theoretical encounter kernel (panel B; here taking a cell radius $r=\SI{1}{\micro\meter}$). The probability decreases moderately with shear rate until a shear rate of $\SI{1e2}{\per\second}$, the optimal shear rate, above which it decreases more rapidly, indicating the increasingly detrimental impact of higher shear on the mating process. Black segments show two different scaling exponents for reference.
\textbf{D}, Encounter duration as a function of shear rate for three different cell radii~(Eq.~\ref{eq:enc_dur}). The theoretical duration over which two cells are close enough to establish a mating pair and can attach to each other, defined as the time required to sweep a volume
equal to the cell volume, is below a millisecond at the highest shear. 
\textbf{E}, Shear force predicted to act on a mating pair as a function of shear rate~(Eq.~\ref{eq:shear_force}) for three different cell radii. The shear force on a mating pair increases linearly with increasing shear rate.
\textbf{F}, Predicted search time of a donor cell for a recipient cells as a function of shear rate~(Eq.~\ref{eq:t_search}, $r=\SI{1}{\micro\meter}$). At low shear rate, a donor finds a recipient within a duration comparable to the experiment duration (30 min), whereas at high shear rate ($\dot{\gamma}=\SI{1e3}{\per\s}$) it takes only a few seconds.
}
    \label{fig:Fig2}
\end{figure*}

We performed experiments to quantify conjugation between \textit{E. coli} donor (TB204) and recipient (NCM $\Delta$mot) cells sharing a conjugative plasmid (either R1-19~\cite{meynell1967mutant} or F~\cite{cavalli1953infective}). Donors and recipients were prepared at the same optical density (OD600 = 0.05 for R1-19; OD600 = 0.20 for F; Methods) in LB medium and then mixed in the 1:1 ratio, with the exact concentrations determined by counting the cells in a hemocytometer under a microscope before each experiment~~(SI Section 2;  Fig.~S1; Table S1). We used low cell concentrations to minimize the chance of non-binary collisions (e.g., between three cells); the concentrations were higher for the F plasmid to compensate for its lower transfer efficiency compared to the R1-19 plasmid. We then exposed the donor-recipient mixture to a constant shear flow in the rheometer for $t_\tn{m}=\SI{30}{\min}$ at a constant temperature of $T=\SI{30}{\degree C}$. We then sampled cells to count transconjugants using the time to threshold method~\cite{bethke2020environmental}, which we extended to include the impact of stochastic cell division (Methods, SI Section 3).  
Finally, we computed the conjugation rate $\eta$ as~\cite{Huisman2022, Kosterlitz2023, Lopatkin2016,bethke2020environmental}
\be
\label{eq:eta}
\eta = \f{c_\tn{T}(t_\tn{m})}{t_\tn{m} c_\tn{D}(0)c_\tn{R}(0)},
\ee
where $c_\tn{D}(0)$ and $c_\tn{R}(0)$ are the initial concentrations of donor and recipient cells, respectively, and $c_\tn{T}(t_\tn{m})$ is the transconjugant endpoint concentration. In $n=4$ experiments, the average initial concentrations were
$c_\tn{D}=\SI{1.49e7}{\per\milli\liter}\pm\SI{2.58e6}{\per\milli\liter}$ and $c_\tn{R}=\SI{1.05e7}{\per\milli\liter}\pm\SI{1.58e6}{\per\milli\liter}$ for R1-19; $c_\tn{D}=\SI{4.93e7}{\per\milli\liter}\pm\SI{1.41e7}{\per\milli\liter}$ and $c_\tn{R}=\SI{4.02e7}{\per\milli\liter}\pm\SI{1.12e7}{\per\milli\liter}$ for F ($\tn{mean} \pm \tn{sd}$).

Our experiments revealed that the conjugation rate is strongly dependent on the shear rate~(Fig.~\ref{fig:Fig1}D). To measure the same sample across a broad range of shear rates, we split the same bacterial culture into aliquots and sequentially exposed them to different shear rates. We changed the shear rate from low to high in eight steps in two series (`up' series) and from high to low in two series (`down' series). Each series was performed on a different day. We found that, as a function of shear rate, the conjugation rate $\eta$ increases from a plateau at low shear rates ($\dot\gamma<\SI{1}{\per\second}$) to a maximum at a critical shear rate of $\dot\gamma=\SI{1e2}{\per\second}$, and then decreases again~(Fig.~\ref{fig:Fig1}D). This behavior is observed for both the up and down series~(shaded lines in Fig.~\ref{fig:Fig1}D, Fig.~S2A,B). Averaging the conjugation rate across the four series showed that the peak conjugation rate was approximately five times higher than the plateau at a low shear rate. The observed conjugation rates for both plasmids were high compared to other plasmids~\cite{Sheppard2020} as expected for these permanently derepressed (i.e., constitutively expressed) conjugative plasmids~\cite{meynell1967mutant,yoshioka1987repressor}. The F plasmid exhibited a lower conjugation rate than R1-19, but scaling the values of the conjugation rate for the F plasmid by a constant factor~($550\times$) showed that it exhibits a very similar response to shear as R1-19.

 For the R1-19 plasmid, we additionally confirmed the five-fold increase in the conjugation rate in separate experiments where the conjugation rate of a sample exposed to a shear rate of $\dot\gamma=\SI{1e2}{\per\second}$ was compared with a simultaneous (rather than sequential) no-flow control~(Figs.~S2C and D; Table S2). To eliminate the possibility that the increase in the conjugation rate is driven not directly by shear but by heterogeneity in cell concentrations, we measured the cell concentration at three different positions (center, midpoint and edge of the cone) by sampling $\SI{50}{\micro\liter}$ after lifting the cone. We did so for two samples, one when the cone was not rotating and one for a sample sheared at $\dot\gamma=\SI{1e2}{\per\second}$~(Fig.~S3A). When the cone was not rotating, cells slowly accumulated near the edge of the sample, likely due to evaporation-driven flow~\cite{sempels2013auto,ruan2023evaporation}. In contrast, cells sheared at $\dot\gamma=\SI{1e2}{\per\second}$ remained more homogeneously distributed in the rheometer (Fig.~S3B). This confirms that the increase in conjugation rate at intermediate values of shear does not result from the creation of local hotspots of high cell concentration. Finally, we carried out a further control experiment to eliminate the possibility that the plasmid transfer occurs via non-contact horizontal gene transfer pathways, e.g., through the shedding of plasmids or extracellular vesicles by donors into the liquid, followed by uptake by recipients (Fig.~S3C).

We next rationalize the impact of the shear rate on the conjugation rate (Fig.~\ref{fig:Fig2}A) using encounter rate theory. We model the encounter rate $E$ between donor cells (concentration $c_\tn{D}(t)$ at time $t$) and recipient cells (concentration $c_\tn{R}(t)$ at time $t$) per unit volume in the rheometer as
\be
\label{eq:enc_rate}
E = \Gamma c_\tn{D}(t) c_\tn{R}(t),
\ee
where $\Gamma$ is the encounter kernel, which represents the relative volume swept per unit time by a pair of colliding cells~\cite{kiorboe2009mechanistic,slomka2023encounter}. $\Gamma$ has the same units as the conjugation rate $\eta$ and explicitly depends on the encounter mechanism. We make the common assumption that shear flow and diffusion generate encounters independently~\cite{burd2009particle}, implying that the total kernel is a sum of two separate kernels
\be
\label{eq:gamma_tot}
\Gamma=\Gamma_\tn{shear}+\Gamma_\tn{diff},
\ee
with~\cite{Smoluchowski1916,burd2009particle}
\bse
\label{eq:gammas}
\be
\label{eq:gamma_shear}
\Gamma_\tn{shear}&=&\f{4}{3}(r_\tn{D}+r_\tn{R})^3\dot\gamma, \\
\label{eq:gamma_diff}
\Gamma_\tn{diff}&=&4\pi(D_\tn{D}+D_\tn{R})(r_\tn{D}+r_\tn{R}),
\ee
\ese
where $r_\tn{D}$ and $r_\tn{R}$ are the equivalent radii of donor and recipient cells, respectively, $D_\tn{D}$ and $D_\tn{R}$ are their diffusion coefficients (due to Brownian motion), and $\dot\gamma$ is the shear rate.  In the following, we set $r_\tn{D}=r_\tn{R}=r$, and consider the range $\SI{0.6}{\micro\meter}<r<\SI{1.3}{\micro\meter}$ to represent the variability of cell volume in cultures grown in rich media~\cite{taheri2015cell}. The diffusion coefficient is computed according to the Stokes-Einstein formula, $D=k_\tn{B}T/(6\pi\mu r)$, for a fluid with dynamic viscosity of water at the experimental temperature $T=\SI{30}{\degree C}$~($\mu=\SI{0.8}{\milli\pascal\second}$); $k_\tn{B}$ is the Boltzmann constant. 

Considering the total encounter kernel~(Eq.~\ref{eq:gamma_tot}) as a function of the shear rate for different cell radii $r$~($=r_\tn{D}=r_\tn{R}$) demonstrates that encounters are driven by diffusion rather than shear at low shear rates (Fig.~\ref{fig:Fig2}B). Because of the different scaling of the diffusive and shear-driven kernels with cell size~(Eq.~\ref{eq:gammas}), the total kernel is independent of cell size in the diffusive limit, but is very sensitive to cell size at high shear rates: changing $r=\SI{0.6}{\micro\meter}$ to $r=\SI{1.3}{\micro\meter}$ increases the encounter kernel ten-fold in the shear-dominated regime. Equating $\Gamma_\tn{diff}=\Gamma_\tn{shear}$ in Eq.~(\ref{eq:gammas}) gives the critical shear rate, above which the contribution of shear overtakes that of diffusion in determining the encounter rate (the arrows in Fig.~\ref{fig:Fig2}B indicate this critical shear rate for different cell radii). The calculated critical shear rate~($\dot\gamma=0.6-\SI{6}{\per\second}$) is a close match to the observed transition from the conjugation plateau to the region in which the conjugation rate increases with increasing shear rate.

 To unravel the effects of different environmental settings on the conjugation rate, Sørensen et al. suggested that the efficiency of plasmid transfer should be reported as the number of transfer events per donor-recipient encounter~\cite{sorensen2005studying}. We can accomplish this because of the combination of experimentally controlled shear flow and encounter rate theory. Specifically, assuming that each encounter between a donor and recipient cell has a probability $p$ of resulting in conjugation ~(Fig.~\ref{fig:Fig1}C), the conjugation rate will be related to the encounter kernel as
\be
\label{eq:eta_p_Gamma}
\eta=p\Gamma.
\ee 
In particular, $p=1$ would mean that every encounter results in conjugation, i.e., the conjugation rate is equal to the encounter rate. Note that $\Gamma$ and $\eta$ have the same units of volume per time, and $p$ is a dimensionless free parameter. More generally, if $p$ was constant (i.e., independent of the encounter process or forces acting on the mating pair), then the conjugation rate would have the same response to shear as the encounter kernel. Another limiting case corresponds to $p$ being proportional to the encounter duration and thus inversely proportional to the encounter kernel. In this case, the increase in $\Gamma$ with shear would be perfectly cancelled by a decrease in $p$, and the conjugation rate $\eta$ would be shear-independent.
Dividing the observed mean conjugation rate $\eta$ by the theoretical encounter kernel $\Gamma$  (Eq.~\ref{eq:gamma_tot} computed for $r=\SI{1}{\micro\meter}$) yields the experimentally observed probability $p$ of conjugation per donor-recipient encounter~(Fig.~\ref{fig:Fig2}C). We find that the per-encounter probability of conjugation $p$ decreases with increasing shear rate, suggesting that higher shear makes it harder for cells to form stable mating pairs. This is in line with previous observations in incubator shakers~\cite{Zhong2010}. Above the observed optimal shear rate ($\dot\gamma\approx\SI{1e2}{\per\second}$), the probability of conjugation per encounter decreases more steeply in relation to the shear rate (cf. black segments in Fig.~\ref{fig:Fig2}C). In this regime, the increase in the number of donor-recipient encounters with increasing shear rate~(Fig.~\ref{fig:Fig2}B) is outweighed by the detrimental impact of shear on the conjugation process (Fig.~\ref{fig:Fig2}C), leading to a decrease in the overall conjugation rate (Fig.~\ref{fig:Fig2}A).

We hypothesize that the detrimental impact of shear at high shear rates arises because cell-cell encounters are too short for cells to attach and form productive mating pairs or because the shear force acting on the two cells becomes so strong that it disrupts the mating process. Since we cannot track individual encounter events in our experimental setup, we instead model the average encounter duration $t_\tn{encounter}$ during which cells have a chance to attach as follows~\cite{Hutchinson2007}
\be
\label{eq:enc_dur}
t_\tn{encounter}=V_\tn{cell}/\Gamma,
\ee
where $V_\tn{cell}=4\pi r^3/3$ is the cell volume, again taken to be the same for donors and recipients~($r_\tn{D}=r_\tn{R}=r$). This encounter duration, shown in Fig.~\ref{fig:Fig2}D for different cell radii, represents the time the cells need to sweep a volume equal to the cell volume. Taking $r=\SI{1}{\micro\meter}$ shows that a donor-recipient encounter lasts approximately 0.3 s at low shear rates~($\dot\gamma=\SI{1e-1}{\per\second}$) and drops to 0.4 ms at the highest shear rate~($\dot\gamma=\SI{1e3}{\per\second}$). Because of the different scaling of the diffusive and shear-driven kernels with cell size, the encounter duration is longer for larger cells in the diffusive limit but independent of cell size at high shear rates. The shorter encounter duration may render the attachment step less likely at high shear. 

Even after successful attachment, the shear force can hamper the plasmid transfer. The pilus takes about twenty seconds~\cite{Goldlust2023} to retract and bring the cells together to form a stable pair, although DNA transfer can also occur while the pilus is extended~\cite{Goldlust2023}. The magnitude of the shear force acting upon two touching cells can be estimated as~\cite{goren1971hydrodynamic,Husband1992}
\be
\label{eq:shear_force}
f_\tn{shear}=6.12\pi\mu r^2 \dot\gamma,
\ee
where $\mu$ is the dynamic viscosity of the liquid. Taking $\mu=\SI{0.8}{\milli\pascal\second}$ for water, we find that the shear force acting on the connection between the two cells exceeds 10 pN at the highest shear rate ($\dot\gamma=\SI{1e3}{\per\second}$), even approaching 100 pN for large cells (Fig.~\ref{fig:Fig2}E).
The polymerization machinery in type IV pili produces forces that can, for brief times, reach 100 pN~\cite{merz2000pilus}. Similarly, force-extension measurements of F pili have shown that forces above 10 pN start to mechanically extend a pilus~\cite{Patkowski2023}. Thus, the decrease in the conjugation rate at the highest shear rate ($\dot\gamma=\SI{1e3}{\per\second}$) may arise because shear forces hamper pilus retraction. Alternatively, even if the pilus manages to retract, the shear forces may still separate the mating cells during the subsequent plasmid transfer stage, which takes at least several minutes~\cite{Babic_Science2008,Goldlust2023}. 

While shear may disrupt conjugation, at the highest shear rates reached in our experiment the shear force did not irreversibly damage the cells or pili. To test this, we performed conjugation experiments with the R1-19 plasmid where we first exposed cells to the highest shear rate used in our experiments ($\dot\gamma=\SI{1e3}{\per\second}$) and then to the optimal shear rate ($\dot\gamma=\SI{1e2}{\per\second}$). We compared the measured conjugation rates with a control experiment in which cells were exposed only to the optimal shear rate~(SI Section 4, Fig.~S4). These experiments showed that cells fully recovered the same maximal conjugation rate at the optimal shear rate despite initial exposure to the highest shear rate. Overall, our observations and calculations suggest that the decrease in the conjugation rate at the highest shear is likely driven not by a cellular stress response but by the mechanical mechanisms of either a decreased rate of pilus attachment due to short encounter durations or the separation of mating pairs by the high shear forces.


\begin{figure*}[t!]
    \centering
    \includegraphics[width=1.0\textwidth]{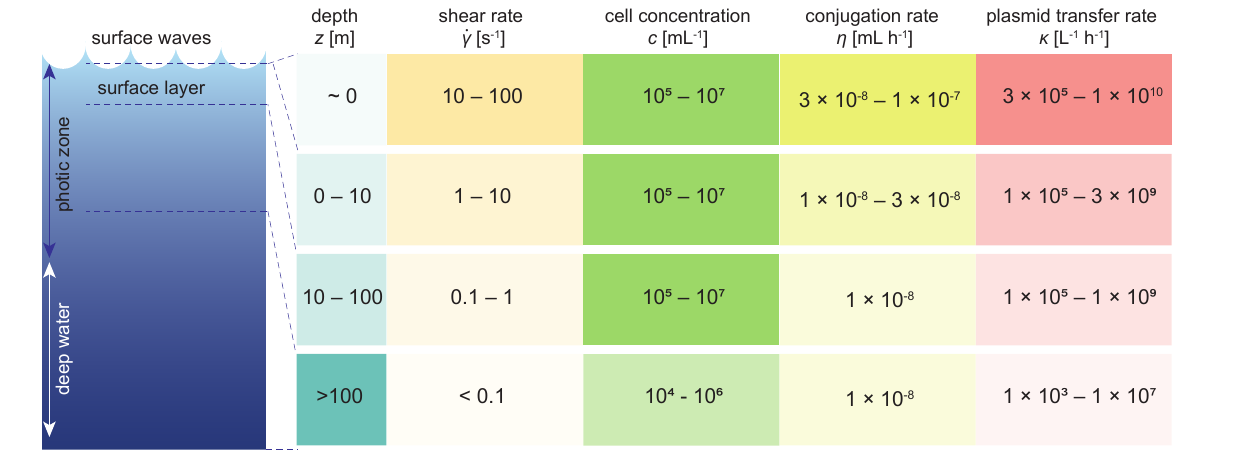}
    \caption{
    \textbf{The surface layer and breaking waves are predicted to represent high-shear-mediated conjugation hotspots in the ocean}. 
    The table gives estimates of the conjugation rate and the number of plasmid transfers per unit volume and unit time at different depths in the ocean. Estimates are based on the experimental relationship between conjugation rate and shear rate reported here (Fig.~\ref{fig:Fig1}D) and observational data of shear rates and cell concentrations at different depths in the ocean (see Eq.~(\ref{eq:plasmid_transfer}) and main text).
    }
    \label{fig:Fig3}
\end{figure*}


The duration of conjugation events, rather than the search time between events, likely limits the measured maximum conjugation rate at intermediate and high shear rates. From the perspective of a single donor cell, it takes on average a search time~\cite{kiorboe2009mechanistic,slomka2023encounter}
\be
\label{eq:t_search}
t_\tn{search}=(\Gamma c_\tn{R})^{-1}
\ee
 to encounter a recipient cell. This time is illustrated in Fig.~\ref{fig:Fig2}F for the recipient cell concentrations used in our experiments and $r=\SI{1}{\micro\meter}$. The search time is of the same order as the experiment duration (30 min) at low shear rates and drops to around one second at the highest shear rates. Thus, over the duration of the experiment, a donor encounters more than a thousand potential recipients at high shear but only, on average, one recipient at low shear. Given that conjugation lasts at least several minutes~\cite{Babic_Science2008,Goldlust2023}, this estimate implies that the observed peak conjugation rate may be limited not by the search time but by the conjugation duration. Namely, donors may find a viable conjugation partner rapidly~($\sim$ 1 min) at the optimal shear rate ($\dot\gamma=\SI{1e2}{\per\second}$) and then spend a much longer period conjugating, and are thus able to repeat the whole cycle at most a handful of times during the experiment, leaving many encounters that occur in the meantime unused. In a more dilute system, or when only a small subpopulation of cells represent viable recipients (e.g., due to phylogenetic distance~\cite{Sheppard2020} or incompatible mating pair stabilization mechanisms~\cite{low2022mating}), cells may be limited by the search time at all shear rates. In such scenarios, we expect that the maximum enhancement in the conjugation rate due to shear over diffusion alone will be even higher than the factor of five observed here. 

The presence of a maximum in the conjugation rate as a function of shear rate is robust to selected perturbations of experimental parameters. So far, we performed the experiments at a constant shear rate (with the cone rotating at a fixed angular speed for a given shear rate), constant temperature ($\SI{30}{\degree C}$), and in the same medium as the growth medium (LB). We varied these parameters to test the robustness of the shear response curve using the R1-19 plasmid (Fig.~S5). We first performed a conjugation experiment in oscillatory shear, where the amplitude and frequency of the oscillations were chosen to match the shear rate magnitude in the constant-shear experiments. This oscillatory shear experiment yielded a response curve that was similar to that observed in the constant-shear experiments (Fig.~S5A). We further performed experiments in constant shear but at different temperatures (27, 30, and $\SI{37}{\degree C}$). Changes in temperature only rescaled the response curve by a multiplicative prefactor without affecting the optimal shear rate value ($\dot\gamma=\SI{1e2}{\per\second}$; Fig.~S5B). Finally, we performed additional conjugation experiments in M9 minimal medium for cells grown in LB medium, to mimic a sudden exposure to a low-nutrient environment. In M9 medium, the shear response curve only decreased by a multiplicative prefactor (Fig.~S5C), a response similar to a decrease in temperature. These observations are consistent with the finding that the shear response curve results from the mechanics of encounters and hydrodynamic forces acting on mating cells.

Relating laboratory measurements of conjugation rates to environmentally relevant conditions is important to understand the constraints on horizontal gene transfer rates in the environment~\cite{Aminov2011,larsson2022antibiotic}. Our experimental conditions most closely mimic those in aquatic environments due to the relatively low cell densities ($\sim\SI{e7}{\per\milli\liter}$) and low viscosity (LB medium). Marine environments host diverse microbiota~\cite{sunagawa2015structure} that experience vastly different levels of turbulence (and thus shear), for example due to decreasing turbulent intensity with depth~\cite{Sutherland2015,Franks2022}. We therefore next leverage our observations to quantify the effect of shear on conjugation rates at different depths in the ocean.

Turbulence in the ocean's surface waters may elevate the conjugation rate in comparison with more quiescent deeper waters, with breaking waves acting as a conjugation hotspot~(Fig.~\ref{fig:Fig3}). 
We predict the conjugation rate in turbulence by connecting the mean encounter rate in turbulence with the mean encounter rate experienced by cells in our experiments as follows. Bacteria are much smaller than the Kolmogorov scale, the smallest scale in a turbulent flow~\cite{thorpe2007introduction}, which in the ocean is typically in the range 1-6 mm~\cite{lazier1989turbulence}. At such small scales, the encounter kernel characterizing cell-cell encounters is~\cite{Saffman1956} 
\be
\label{eq:Gamma_turb}
\Gamma_\tn{turb}=1.3(r_\tn{D}+r_\tn{R})^3 \sqrt{\epsilon/\nu},
\ee
where $\epsilon$ is the kinetic energy dissipation rate and $\nu$ is the kinematic viscosity of water. The kernel in Eq.~(\ref{eq:Gamma_turb}) is identical with the kernel $\Gamma_\tn{shear}$ in Eq.~(\ref{eq:gamma_shear}) to within 3\%, when one uses the turbulent shear rate $\dot\gamma= \sqrt{\epsilon/\nu}$. Based on this similarity between the kernels, we can assume the probability of conjugation per encounter $p$ to have the same dependence on shear as in our experiments, which enables us to translate environmental shear rates into conjugation rates on the basis of our observed relationship~(Fig.~\ref{fig:Fig1}D).  

Below an ocean depth of ten meters, turbulent shear rates are typically low ($\dot\gamma< \SI{1}{\per\second}$)~\cite{Franks2022}, which corresponds to the diffusion-dominated regime with cell-cell encounters driven by Brownian motion (the low-shear plateau in our experiments). Hence, to model a plasmid spreading in deeper waters (>10 m), we take the conjugation rate $\eta=\SI{1e-8}{\milli\liter\per\hour}$ for the  R1-19 plasmid~(the plateau value in Fig.~\ref{fig:Fig1}D). Conversely, the shear rate in the ocean's upper ten meters is in the range $\dot\gamma= \SI{1e0}{\per\second}\tn{ to }\SI{1e1}{\per\second}$, and breaking waves can generate $\dot\gamma =\SI{1e2}{\per\second}$~\cite{Sutherland2015}. This approximately corresponds to values of $\eta=\SI{1e-8}{\milli\liter\per\hour}\tn{ to }\SI{3e-8}{\milli\liter\per\hour}$ in the upper ten meters and extremes of $\eta=\SI{1e-7}{\milli\liter\per\hour}$ in breaking waves (the peak value in Fig.~\ref{fig:Fig1}D). The highest shear rate in our experiments ($\dot\gamma \approx\SI{1e3}{\per\second}$) is rarely, if ever, observed in the ocean, implying that ocean turbulence is too weak to disrupt conjugation mechanically. 

Based on a global survey of microbial abundances, we take cell concentrations in the range $c= \SI{1e5}{\per\milli\liter}\tn{ to }\SI{1e7}{\per\milli\liter}$ in the upper 100 m and $c= \SI{1e4}{\per\milli\liter}\tn{ to }\SI{1e6}{\per\milli\liter}$ at depths below 100 m~\cite{Wigington2016}. Combining these estimates, we compute the plasmid transfer rate $\kappa$ per liter per hour as a function of depth $z$ as follows~
\be
\label{eq:plasmid_transfer}
\kappa(z) = \eta[\dot\gamma(z)]c^2(z),
\ee 
where we assumed $c_\tn{D}=c_\tn{R}=c$. Fig.~\ref{fig:Fig3} summarizes this prediction and suggests that, below ten meters, conjugation is driven primarily by diffusion. By contrast, breaking waves could act as conjugation hotspots, based on the combination of high cell concentrations and a strong enhancement of the conjugation rate by turbulent shear. Next, we compare this maximum conjugation rate with that of spontaneous mutations.

Conjugation in the turbulent surface ocean may generate genetic variation faster than spontaneous mutation. Because conjugation rates, in general, depend on phylogeny~\cite{Sheppard2020,Castaneda-Barba_NMR2024}, we focus on the marine \textit{Roseobacter} clade. \textit{Roseobacter} represents up to 20\% ($\beta=0.2$) of bacterial cells in some coastal ecosystems and 3 to 5\% ($\beta=0.03-0.05$) of bacterial cells in open ocean surface waters~\cite{moran2007ecological}, and can share mobilizable marine plasmids~\cite{petersen2017plasmid,Petersen2019}. If only a fraction $\beta$ of cells in the community (i.e., \textit{Roseobacters}) participates in conjugation, then the estimates in Fig.~\ref{fig:Fig3} decrease by a factor of $\beta^2$. To compare the plasmid transfer rate with the spontaneous mutation rate, we assume cells divide once per day and mutate spontaneously at a rate of $2.5\times 10^{-3}$ mutations per genome per replication~\cite{drake1998rates}. Considering conditions with a large total cell concentration ($\SI{1e6}{\per\milli\liter}$), in which \textit{Roseobacters} represent a large fraction of the population ($\beta=0.1$), and taking $\eta=\SI{1e-7}{\milli\liter\per\hour}$ for the conjugation rate in the surface ocean~(Fig.~\ref{fig:Fig3}), we find that the plasmid transfer rate is two orders of magnitude higher than the spontaneous mutation rate (see SI Section 5 for a detailed calculation and Fig.~S6).
These estimates indicate that conjugation in the turbulent surface ocean has the potential to generate faster genetic adaptation in marine bacteria than spontaneous mutations.

\subsection*{Discussion}

We have presented the results of bacterial conjugation experiments under controlled shear conditions. While previous experiments suggested an important role for fluid shear in controlling conjugation efficiency~\cite{Clarke2008,Zhong2010,Patkowski2023}, they were based on conventional laboratory shakers~\cite{Zhong2010,Patkowski2023}. Shakers generate flows characterized by a broad distribution of shear rates that depends on the flask shape and sample volume~\cite{Li2013}. Poorly constrained shear rates, in turn, prevent a quantitative and mechanistic understanding of the opposing impacts of shear on conjugation, which can act to increase the conjugation rate by increasing cell-cell encounters but can also disrupt the mating pairs~\cite{Zhong2010}. We have overcome this limitation by quantifying conjugation rates within a cone-and-plate rheometer, which creates a controlled shear flow and, thus, a controlled donor-recipient encounter rate. We discovered that the conjugation rate increases with shear until it peaks at an optimal shear rate, reaching a value five-fold higher
than the baseline conjugation rate set by diffusion-driven encounters. The precise control over cell-cell encounters in the rheometer combined with encounter rate theory enabled us to map out the conjugation rate as a function of depth in a marine environment. Consequently, we predicted that waves at the ocean surface could act as
hotspots of conjugation by generating high enough shear to increase cell-cell encounters without impairing conjugation and that conjugation can drive faster genetic adaptation than spontaneous mutations. 

The dependence of the conjugation rate on shear for the two plasmids we tested, R1-19 and F, was very similar upon a simple rescaling (Fig.~\ref{fig:Fig1}D), suggesting a shear-independent factor (e.g., lower number of pili or less successful establishment in a new cell) that accounts for the lower conjugation probability per donor-recipient encounter for the F plasmid. It remains to be seen whether this shear dependence holds universally across different bacterial species and plasmid types. In particular, while plasmids carrying type-IV secretion systems are abundantly present among Roseobacters~\cite{Rapp2024}, their mechanical properties remain unknown. Plasmids used in this study belong to the IncF incompatibility class, which produce long, flexible, and retractable pili~\cite{bradley1980specification,arutyunov2013f}. Plasmids in the IncW or IncP incompatibility classes produce short, rigid pili, conjugate poorly in liquids, and require surfaces to support mating~\cite{bradley1980specification}. For such plasmids, we expect that the optimal conjugation rate occurs at much lower shear rates than for plasmids in the IncF class. Conversely, the stickiness mediated by the presence of compatible mating pair stabilization protein complexes~\cite{low2022mating} may, in general, enable mating pairs to withstand higher shear. Numerical models of sticky cells connected by retracting, elastic pili, and moving in a shear flow would help identify critical shear rates in relation to the mechanical properties of pili.

Motility can further increase the donor-recipient encounter rate. In our experiments, we deliberately isolated the role of shear from that of motility by using a nonmotile recipient and a donor strain in which only a small subpopulation was moderately motile after centrifugation~(Fig.~S7A,B). However, the encounter kernel $\Gamma_\tn{swim}=4/3\pi(r_\tn{D}+r_\tn{R})^2 U_\tn{swim}$ characterizing encounters between cells swimming in random directions~\cite{slomka2023encounter} implies that a population of cells swimming at speed $U_\tn{swim}=\SI{20}{\micro\meter\per\second}$, typical for marine bacteria~\cite{johansen2002variability}, generates as many cell-cell encounters as the shear rate $\dot\gamma=\SI{10}{\per\second}$, corresponding to strong turbulence. Because motility is an independent encounter mechanism, it may thus further increase the encounter rate between cells on top of the diffusion-driven and shear-driven encounters. Conversely, motility likely decreases encounter durations. Up to 10\% of bacteria can be motile in coastal waters~\cite{mitchell1995long}, raising the possibility that motility could be an important phenotype affecting conjugation in bacterial populations.

Apart from using model organisms and plasmids, our extrapolation to the environment is based on several simplifying assumptions. While our observation that conjugation peaks at the optimal shear rate is robust to selected perturbations of the experimental parameters~(Fig.~S5), including oscillatory shear, turbulence in aquatic environments~\cite{Franks2022} or gut mixing~\cite{schutt2022simulating} can exhibit high levels of intermittency. Consequently, a more accurate identification of conjugation hotspots may require resolving stirring in natural environments at high spatial and temporal resolution, as well as integrating the effect of stirring on conjugation over the history of shear rate experienced by mating pairs. Additionally, aggregates are abundant in aquatic ecosystems~\cite{burd2009particle} and may provide different microenvironments for conjugation. Cells in our experiments did not aggregate (Fig.~S7C), validating the description of conjugation based on binary collisions (Eq.~\ref{eq:eta}). However, when aggregates are present, aggregation formation~\cite{Sinzato2025} and the dynamics of conjugation within aggregates~\cite{djermoun2025biofilm} must be accounted for when predicting conjugation rates. 

 Constraining horizontal gene transfer rates in natural settings is important for identifying bottlenecks and drivers of microbial evolution, including the spread of antibiotic resistance~\cite{larsson2022antibiotic,Castaneda-Barba_NMR2024,shepherd2024ecological}.
 Preferential flow paths in porous media (e.g., in soils) can generate shear rates that span several orders of magnitude and cover the whole range of values we investigated ~\cite{kurz2022competition}. Gut peristalsis can generate shear rates up to $\dot{\gamma}\approx\SI{1e1}{\per\s}$~\cite{schutt2022simulating}. Given that the high viscosity of the gut digesta likely suppresses diffusion while the host immune system can quench motility~\cite{cullender2013innate}, shear-driven encounters may be the dominant physical mechanism of conjugation in the gut of hosts. Overall, by increasing the rate of shear-driven cell-cell encounters, we predict that environmental flows generate bacterial conjugation hotspots in the environment.

\newpage
\textbf{Acknowledgements}

We gratefully acknowledge funding from the National Science Foundation under PHY-2309135~(KITP HGT24) to M.Z, J.S.H. and J.S.; a Human Frontier Science Program (HFSP) Postdoctoral Fellowship LT0045/2023-L to J.S.H.; a Simons Foundation Pivot Fellowship to N.B. and R.S.; a Gordon and Betty Moore Foundation Symbiosis in Aquatic Systems Initiative Investigator Award (GBMF9197), the Simons Foundation through the Principles of Microbial Ecosystems (PriME) collaboration (grant 542395FY22), Swiss National Science Foundation grant 205321\texttt{\char`_}207488, Swiss National Science Foundation Sinergia grant CRSII5-186422, and the Swiss National Science Foundation, National Centre of Competence in Research (NCCR) Microbiomes (Nos. 51NF40\texttt{\char`_}180575 and 51NF40\texttt{\char`_}225148) to R.S.; and a Swiss National Science Foundation Ambizione grant no. PZ00P2\texttt{\char`_}202188 to J.S.

We thank Deepthi Vinod and David Johnson for providing us with the donor strain with the two plasmids and Roberto Pioli for providing us with the recipient strain. 
We thank Yao Zhou and Franciszek Myck for assistance, and Uria Alcolombri and Sebastian Bonhoeffer for discussions.

Figures were partly generated using Servier Medical Art, provided by Servier, licensed under a Creative Commons Attribution 3.0
unported license.

\textbf{Data and materials availability} 
All data are available in the manuscript, the supplementary material or deposited at ETH Research Collection~\cite{Slomka_ETH_RC_2025}.

\textbf{Author contributions}
M.Z. and J.S. designed research; M.Z. and J.S. performed research; M.Z., J.S.H., and J.S. analyzed data; M.Z., N.B., and J.S. performed statistical analysis and analyzed the stochastic time to threshold method;  J.S.H. had input on the design of the research; and M.Z., J.S.H., N.B., R.S., and J.S. wrote the paper.

\textbf{Competing interests} 
The authors declare no competing interests.

\textbf{Materials and Methods}

\textit{Bacterial strains} As plasmid donor, we used \textit{Escherichia coli} strain TB204 $\Delta$trpC-GFP, a derivative of strain MG1655 made auxotroph for tryptophan and fluorescently labeled with the fluorophore sfGFP on the chromosome~\cite{dal2020short}. In our experiments, the strain carried one of two plasmids. The first is the R1-19 plasmid, a self-mobilizable, derepressed, and upregulated variant of the R1 plasmid that also contains a resistance gene against chloramphenicol~\cite{meynell1967mutant}. The second plasmid is a variant of the F-plasmid~\cite{cavalli1953infective} containing an additional resistance gene against tetracycline. As recipient strain, we used a $\Delta$motA mutant of strain NCM3722~\cite{taheri2015cell}, which lacks flagella and is non-motile. Furthermore, we used the mini-Tn7 insertion~\cite{choi2006mini} to chromosomally tag this strain with the fluorophore dsRedExpress and a resistance gene against gentamicin. The insertion (pUC18T-mini-Tn7T-Gm-dsRedExpress; Addgene \#65032) and helper plasmid (pTNS1; Addgene \#64967) were gifts from Herbert Schweizer. 

\textit{Media preparation} Standard Luria broth (LB; Difco\textsuperscript{TM} LB Broth, Miller (Luria-Bertani) from Becton, Dickinson and Company) was used as the medium for cell culturing, conjugation experiments, and transconjugant selection. Antibiotics were added to that broth during culturing and selection (as specified below). For the experiment in the standard M9 medium (Fig.~S5C), we prepared twofold diluted M9 Minimal Salts 2X (Gibco), which was enriched with 0.4\% glucose (filter-sterilized), 2 mM magnesium sulfate, and 0.1 mM calcium chloride, and then autoclaved.

\textit{Culture preparation}
Prior to experiments, cultures of one donor strain and the recipient strain were grown overnight with shaking (200 rpm) at $\SI{30}{\degree C}$ in LB medium containing one of the following antibiotics: $\SI{50}{\micro\gram\per\milli\liter}$ chloramphenicol (R1-19), $\SI{10}{\micro\gram\per\milli\liter}$ tetracycline (F), or $\SI{60}{\micro\gram\per\milli\liter}$ gentamicin (recipient). The following day, samples of each culture were diluted 100x in fresh LB containing the same antibiotics at the same concentrations and further incubated at $\SI{30}{\degree C}$ with shaking at 200 rpm for 2 h to ensure cells were in exponential growth phase; typical OD600 values were in the range of 0.05-0.2 at this point. Thereafter, the cultures were centrifuged (2000~rcf at $\SI{25}{\degree C}$ for 10 min), and after removing the supernatant, the cells were resuspended in LB free of antibiotics. We then diluted the suspensions to specific OD600 values (0.05 for R1-19 donors and recipients, 0.20 for F donors and recipients). These cell suspensions were then split into eight pairs of donor and recipient aliquots, and each pair assayed at a different shear rate in the rheometer on the same day. To determine the concentrations of donors and recipients, samples of $\SI{15}{\micro\liter}$ were taken from the aliquots prepared for the first and last assay of the day, and the cell density was measured using a hemocytometer immediately after each assay started (Bright-Line, Hausser Scientific; SI Section 2; Figs.~S1A--C). Until used in experiments, aliquots were stored at $\SI{4}{\degree C}$ in a fridge. We observed no significant cell growth in these aliquots during this storage (Fig.~S1D) and no significant impact of the duration spent at $\SI{4}{\degree C}$ on the shear-induced enhancement of the conjugation rate~(Fig.~S2D).

\textit{Shear-flow driven conjugation assay}
For each assay, one aliquot of donors and one of recipients (equal volumes) were mixed and vortexed for 10 s to create a homogeneously mixed donor-recipient cell suspension. Of this mixture, 1.2 mL was loaded onto the rheometer (Kinexus lab+, \#KNX2112, NETZSCH) after pre-warming it to $\SI{30}{\degree C}$ using a Peltier plate temperature control module (\#KNX2001-E, NETZSCH) and heat exchanger (\#KNX2500, NETZSCH). Then the upper, cone-shaped geometry ($\SI{4}{\degree}$ angle; stainless steel; \#KNX2036, NETZSCH) was lowered onto the liquid, the moisture trap closed around it and the liquid stirred to generate the desired shear rate for $t_\tn{m}=$ 30 min at  $\SI{30}{\degree C}$.
Each of the eight individual assays was exposed to a different shear rate, with assays ordered from low ($\dot{\gamma}=\SI{1e-1}{\per\s}$) to high ($\dot{\gamma}=\SI{1e3}{\per\s}$) or from high to low on any given day. For each plasmid, we repeated each experiment four times on different days, two in ascending order (up series) and two in descending order (down series). We determined the optimal conjugation assay duration ($t_\tn{m}=$ 30 min) in a separate experiment where we took samples from the running rheometer every 15 min over one hour; 30 min was the shorest time above which the transconjugant concentration increased linearly with time (Fig.~S1E). Our approach to use the rheometer is inspired by mixers developed to study phytoplankton coagulation~\cite{kiorboe1990coagulation} and fertilization in sea urchins~\cite{mead1995effects} based on an oscillating stirring shaft \cite{kiorboe1990coagulation} or the Couette cell ~\cite{mead1995effects}, though these were developed for larger cells (>$\SI{10}{\micro\meter}$) and sample volumes (>$\SI{10}{\milli\liter}$). 

\textit{Transconjugant quantification via time to threshold}
After experiments, the cone-shaped upper geometry was gently raised, the sample mixed by aspiration with a pipette five times, and a subsample taken from the cell suspension in the rheometer ($\SI{200}{\micro\liter}$ if R1-19 was the plasmid; $\SI{400}{\micro\liter}$ if F was the plasmid). This sample was diluted (mixing by vortexing for 5 s) in a selective medium of LB cooled to $\SI{4}{\degree C}$ containing $\SI{60}{\micro\gram\per\milli\liter}$ gentamicin and either $\SI{50}{\micro\gram\per\milli\liter}$ chloramphenicol (R1-19) or $\SI{10}{\micro\gram\per\milli\liter}$ tetracycline (F), in a volume of either $\SI{1800}{\micro\liter}$ (R1-19, 10x dilution) or $\SI{1600}{\micro\liter}$ (F, 5x dilution). The dilution of the sample was performed to reduce the encounter rate due to diffusion and thus the number of possible post-experiment conjugation events, while the exposure to cold temperature was used to suppress the growth of transconjugants, the only cell type able to grow in this medium with both antibiotics present. These diluted samples were stored at $\SI{4}{\degree C}$ until the last experiment of that day's series had been performed (maximum period approximately 5 h).
To quantify the transconjugants, aliquots of $\SI{200}{\micro\liter}$ of these diluted samples were loaded onto a 96-well-plate, with 9 technical replicates created from the sample of each assay (see Fig.~S8A for a typical plate organization). In addition, 16 wells on the plate were used to perform positive and negative controls, including the quantification of the number of additional transconjugants formed in the plate reader after the actual assay. The concentrations of these post-assay transconjugants were orders of magnitude lower than the concentrations observed in the assays (Fig.~S8B). The 96-well-plate was then put into an automated plate reader (BioTek Synergy H1 and BioTek Synergy HTX, Agilent), where the transconjugant cells could grow at a temperature of \SI{37}{\degree C} for the next 24-72 h while the instrument measured the OD600 values of all wells in intervals of 5 min.
The resulting growth curves (Fig.~S8B) were used to estimate the initial concentrations of transconjugants for each technical replicate using the time to threshold method  ~\cite{bethke2020environmental}, and our own calibration curve (Fig.~S1F). Chromosomal tagging of cells with fluorescence markers served as an additional control for selection in a selected experiment~(Fig.~S9). Furthermore, we extended the time to threshold method to account for stochastic effects induced by cell division during the outgrowth stage (SI Section 3; Fig.~S10). Tables S1 and S2 provide a summary of measured cell concentrations in the experiments.

\textit{Statistical analysis}
Detailed information about the analysis of the variability in the technical and biological replicates, error propagation, and statistical tests can be found in SI Section 6. Briefly, to calculate the conjugation rate at a given shear rate, we used Eq.~(\ref{eq:eta}). The average initial concentrations of donors $c_\tn{D}(0)$ and recipients $c_\tn{R}(0)$ were determined by pooling cell counts from the first and last assay of the day (SI Section 2). For the endpoint transconjugant concentration $c_\tn{T}(t_\tn{m})$, we used the average transconjugant concentration measured in the different technical replicates of an individual conjugation assay at a given shear rate. Each technical replicate corresponded to a cell concentration obtained by converting OD600 readouts from individual wells in the well-plate to cell concentrations using the calibration curve (the time to threshold method; SI Section 2). We then used these average values of $c_\tn{D}(0)$, $c_\tn{R}(0)$ and $c_\tn{T}(t_\tn{m})$ (Tables S1 and S2) in Eq.~(\ref{eq:eta}) to compute the conjugation rate of a single biological replicate. We then computed the final value of the conjugation rate by averaging over the biological replicates. For the sequential up/down series experiments and simultaneous shear vs. no shear experiments, we tested the statistical significance of the impact of shear on the conjugation rate using the one-sided two-sample Kolmogorov-Smirnov test. As samples, we used different biological replicates performed on different days. We compared the values of the conjugation rate at the optimal shear rate $\dot\gamma=\SI{1e2}{\per\second}$ with the values on the plateau (or without shear). For both types of experiments and both plasmids, we obtained $p<0.01$.

\onecolumngrid
\newpage

\renewcommand{\theequation}{S\arabic{equation}}
\renewcommand{\thetable}{S\arabic{table}}
\setcounter{equation}{0}
\renewcommand{\thefigure}{S\arabic{figure}}
\setcounter{figure}{0}
\renewcommand{\thesection}{\arabic{section}}
\setcounter{section}{0}

\section*{Supplementary Information}
This Section includes:
\begin{itemize}
  \item Supporting text
  \item Figs. S1 to S10
  \item Tables S1 to S2
  \item References
\end{itemize}

\newpage

\section{Secondary and turbulent flows inside the rheometer}
The flow between a shallow rotating cone and a stationary plate has been investigated in detail in the past, both numerically~\cite{fewell1977secondary,oza2021dynamics} and experimentally~\cite{sdougos1984secondary}. At low Reynolds numbers, upon setting the angular speed of the cone to the desired value $\omega$, the fluid flow reaches a steady state within the characteristic time given by~\cite{oza2021dynamics}
\be
t_\tn{r}\approx (r\alpha)^2/(\pi^2\nu),
\ee
where $r$ is the radius of the cone, $\alpha$ is the cone angle and $\nu$ is the kinematic viscosity of water. In our case  $\alpha=4^{\circ}$, $r=\SI{2}{\centi\meter}$, and the kinematic viscosity of water at the experimental temperature $T=\SI{30}{\degree C}$ is $\nu=\SI{8e-7}{\meter\squared\per\second}$, which yields the fast relaxation timescale $t_\tn{r}\approx \SI{0.25}{\second}$, much shorter than the duration of the experiment (30 min). The nature of the flow generated depends on the magnitude of the angular speed $\omega$. At low $\omega$, the flow is laminar, with the flow streamlines confined to the azimuthal direction, and the shear profile is uniform with the shear rate magnitude given by the baseline shear rate $\dot\gamma_\tn{b}=\omega/\alpha$~(see Eq.~(5) in~\cite{fewell1977secondary}). As the rotation speed increases, secondary flows develop due to the centrifugal force in the radial and polar directions~\cite{fewell1977secondary}, and the flow becomes turbulent at high $\omega$~\cite{sdougos1984secondary}. The transitions between the different regimes are captured by a single parameter $\tilde R$~\cite{sdougos1984secondary}
\be
\tilde R = r^2\omega \alpha^2/(12\nu).
\ee
Specifically, the flow is laminar for $\tilde R <0.5$ and becomes turbulent for $\tilde R>4$~\cite{sdougos1984secondary}, which in our experiments corresponds to baseline shear rates $\dot\gamma_\tn{b}\approx\SI{35}{\per\second}$ and $\dot\gamma_\tn{b}\approx\SI{300}{\per\second}$. Thus, for shear rates smaller than $\dot\gamma_\tn{b}\approx\SI{35}{\per\second}$, the flow shear profile is well approximated by the baseline uniform shear rate $\dot\gamma_\tn{b}$. For higher shear rates, additional shear occurs in the rheometer due to secondary flows and, eventually, turbulence. The secondary flows below the turbulent threshold have been studied numerically by Fewell and Hellums~\cite{fewell1977secondary}, who estimated that the magnitude of the effects of the secondary flow on the deformation rate is captured by the following dimensionless number  
\be
\label{eq:Dfewell}
D = 0.1 \tn{Re} \alpha^2(1-\alpha^2),
\ee
where $D$ denotes the maximum magnitude of the radial-polar component of the rate of deformation tensor normalized by the baseline rate and $\tn{Re}=(r/\cos\alpha)^2\omega/\nu$ is the Reynolds number. Thus, $D=0$ for no secondary flows and $D>0$ measures the additional shear rate in the rheometer due to secondary flows as a fraction of the baseline shear rate $\dot\gamma_\tn{b}$. In our system, $D\approx 1$ for $\dot\gamma_\tn{b}\approx\SI{60}{\per\second}$, implying that the additional shear rate due to the secondary flow becomes comparable with the baseline shear; this additional shear is typically localized at the edge of the cone (see. Fig.~8 in~\cite{fewell1977secondary}). For higher shear rates, Eq.~(\ref{eq:Dfewell}) is inaccurate. As an alternative way to quantify the mean additional shear rate induced by the secondary flows or turbulence on top of the baseline shear rate, we balance the power input into the fluid with the kinetic energy dissipation due to fluid viscosity. Previous measurement determined the following empirical expression for the torque needed to sustain the cone's rotation~\cite{sdougos1984secondary}
\be
\label{eq:torque}
T/T_\tn{b}=1+1.29\frac{\tilde R^{3/2}}{3.5+\tilde R},
\ee
where $T_\tn{b}=2\pi\mu\omega r^3/(3\alpha)=2\pi\mu\dot\gamma_\tn{b} r^3/3$ is the torque for the baseline azimuthal flow and $\mu$ is the dynamic viscosity. Eq.~(\ref{eq:torque}) thus measures the additional torque generated by the secondary or turbulent flows as a fraction of the torque generated by the baseline flow. The power generated by the torque is $P=T\omega$ and given that the sample volume is $V=2\pi r^3 \alpha/3$, the power input per unit mass is
\be
\label{eq:power}
P/(V\rho)=\nu\dot\gamma_\tn{b}^2\Big(1+1.29\frac{\tilde R^{3/2}}{3.5+\tilde R}\Big),
\ee
where we used the relation between the kinematic and dynamic viscosities $\nu=\mu/\rho$, and $\rho$ is the fluid density. Eq.~(\ref{eq:power}) shows that the baseline flow dissipates energy per unit mass with the rate $\nu \dot\gamma_\tn{b}^2$. The secondary or turbulent flow must, therefore, provide the additional shear to close the energy balance in the system. We can thus estimate the mean additional shear rate $\dot\gamma_\tn{a}$ as
\be
\label{eq:power2}
P/(V\rho)=\nu\dot\gamma_\tn{b}^2+\nu\dot\gamma_\tn{a}^2,
\ee
which gives
\be
\label{eq:shear_additional}
\dot\gamma_\tn{a}=\dot\gamma_\tn{b}\Big(1.29\frac{\tilde R^{3/2}}{3.5+\tilde R}\Big)^{1/2}.
\ee
For the baseline shear rate $\dot\gamma_\tn{b}=\SI{1e2}{\per\second}$, Eq.~(\ref{eq:shear_additional}) predicts $\gamma_\tn{a}\approx 0.66\gamma_\tn{b}$. For the highest baseline shear rate investigated in our work ($\dot\gamma_\tn{b}=\SI{1e3}{\per\second}$), $\tilde R\approx 14$, which gives $\gamma_\tn{a}\approx 2\gamma_\tn{b}$. At the highest shear, the turbulent regions occupy around half of the sample volume near the cone's edge~(see Fig. 7 in~\cite{sdougos1984secondary}). We used Eq.~(\ref{eq:shear_additional}) as the horizontal error bars in the main text (Fig.~1D and Fig.~2A) and Figs.~\ref{fig:SI_Up_Down_ramp}A and B. Specifically, the one-sided horizontal error bars indicate the intervals $(\dot\gamma_\tn{b},\dot\gamma_\tn{b}+\dot\gamma_\tn{a})$.

\section{Measuring concentrations of donors, recipients and transconjugants}
In this Section, we give further details on the methods we used to count cells. See Tables S1 and S2 for a summary of the measured cell concentrations.

\subsection*{Initial donor and recipient concentrations} 
\begin{figure*}[h!]
    \centering
    \includegraphics[width=1.0\textwidth]
    {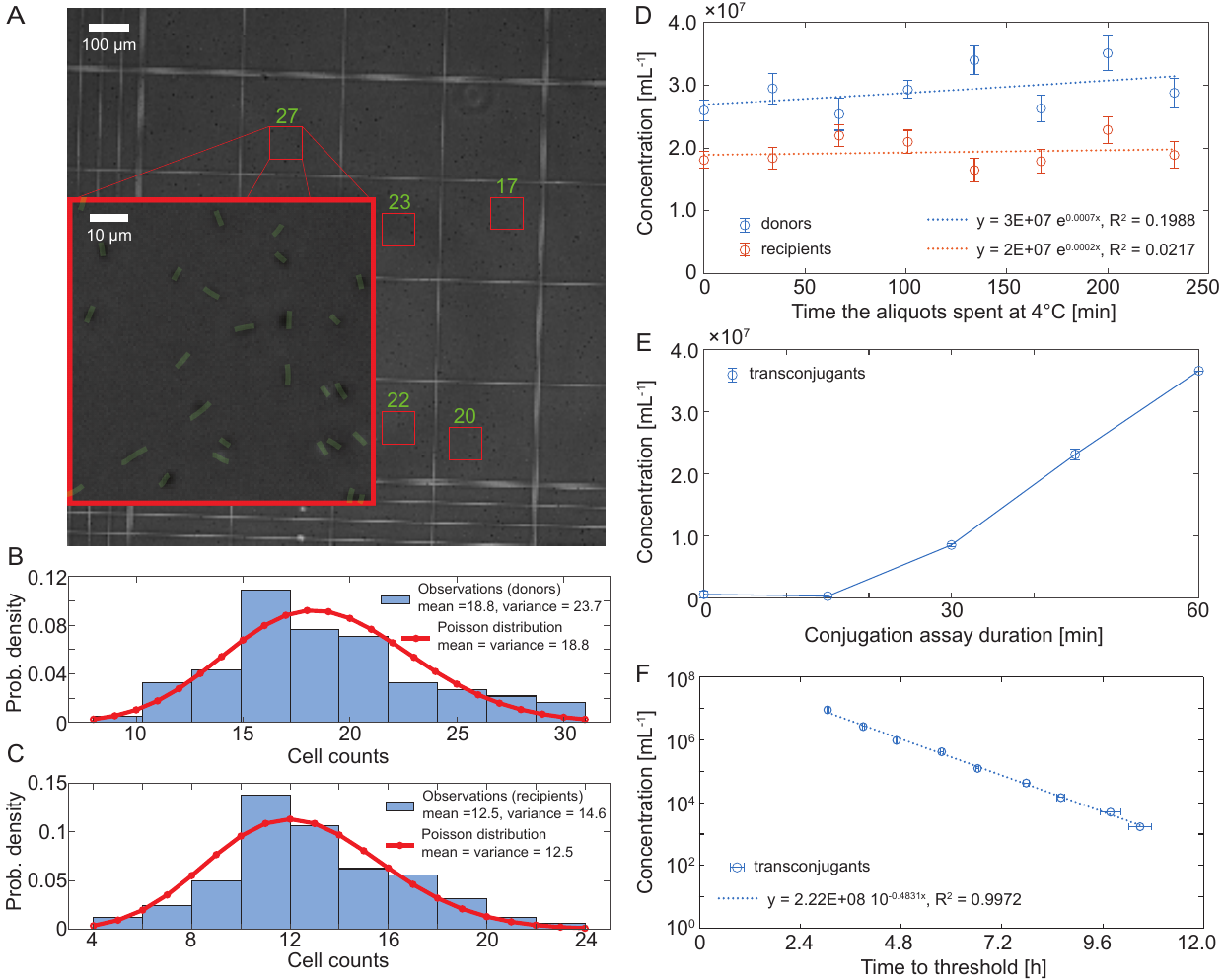}
    \caption{\textbf{Measuring the concentrations of donors, recipients, and transconjugants, and setting the conjugation assay duration.} 
    \textbf{A}, 
    Example of cell counting in the hemocytometer under a brightfield microscope at 10x magnification (SI Section~2).
    \textbf{B, C}, The cell counts in the hemocytometer are approximately Poisson-distributed. The data shows the histogram for R1-19 donors (B) and recipients (C) and the corresponding Poisson distributions with the means equal to the estimated means; the data is pooled from eight separate measurements performed at different times (same data as in panel D).
    \textbf{D}, Concentrations of R1-19 donors (blue) and recipients (orange) in eight pairs of donor-recipient aliquots that spent different amounts of time at $\SI{4}{\degree C}$ before being used in the conjugation assay. 
    The data shows the concentrations determined in the hemocytometer immediately after the respective conjugation assay was started in the rheometer in a single experiment. Each error bar indicates the corresponding standard error of the mean, obtained from the cell counts from the ten evaluated frames in the hemocytometer. The data shows that the cells did not grow while stored at $\SI{4}{\degree C}$ over the course of the experiment.
    \textbf{E}
     To determine optimal conjugation assay duration, we measured the transconjugant concentration as a function of time in a single conjugation assay with the R1-19 plasmid. We exposed a donor-recipient mixture (prepared at concentrations equivalent to an OD600 of 0.05) to a constant shear rate of $\SI{1e2}{\per\second}$, for the total duration of $\SI{1}{\hour}$. Every $\SI{15}{\minute}$, we sampled $\SI{50}{\micro\liter}$ from the running rheometer with a micropipette. This sample was diluted 40x and each dilute loaded into 4 wells (technical replicates) of a 96-well-plate for transconjugant concentration determination using the time to threshold method. The data shows the obtained transconjugant concentrations as a function of the time into the experiment, at which the sample was taken. The error bars indicate the standard error of the mean for these values. At around 30 min, the transconjugant yield has established itself to increase linearly with time, which we chose as the conjugation assay duration in our experiments, except for experiments in Figs.~S2C,D and S3A,B where the duration was 45 min.
    \textbf{F}, The calibration curve of the time to threshold method used to translate the growth times observed in the plate reader~(Fig.~\ref{fig:SI_GrowthCurves}B) into concentrations of transconjugant cells (SI Section~2). The error bars indicate the standard error of the mean observed for the individual threshold times.
    }
    \label{fig:SI_cellcount}
\end{figure*}

In order to get accurate initial concentrations of both donor and recipient cells in our experiments, we counted them directly in a hemocytometer~\cite{Stoddart2011}~(Fig.~\ref{fig:SI_cellcount}A--C). In each series of eight conjugation assays performed on a specific day, a \SI{15}{\micro\liter} sample from the first and last aliquot of the donors and of the recipients was used to make this count. This was done immediately after the respective assay was started to keep the cell concentration in the sample as close to the actual cell concentration used in the assay as possible. Under a brightfield microscope at 10x magnification, images of all four quadrants of the hemocytometer were taken (Fig.~\ref{fig:SI_cellcount}A). Each of these quadrants is divided into 16 squares of side length \SI{250}{\micro\meter}. Out of this total of 64 squares, 10 were selected randomly using a random number generator and into each of these selected squares a quadratic frame of side length $\SI{80.1}{\micro\meter}$ was randomly placed. Then, the number of cells inside these frames was counted by eye. The obtained cell counts follow approximately the Poisson distribution~(Fig.~\ref{fig:SI_cellcount}B and C). The ratios of sample variance to sample mean of 1.3 (donor population) and 1.2 (receiver population) do not suggest significant overdispersion. Because the Poisson distribution describes the number of points randomly and independently distributed in a volume, the fact that the observed counts follow the Poisson distribution validates our procedure of estimating the cell concentrations through random subsampling. With the height inside the hemocytometer being $\SI{100}{\micro\meter}$ (and the volume enclosed by each frame thusly being $\SI{6.416e-7}{\milli\liter}$), this allowed us to determine a value for the cell concentration of this sample from each of the ten frames. Since during the storage at \SI{4}{\degree C} there is no effective growth within the aliquots (Fig.~\ref{fig:SI_cellcount}D), we pooled together the obtained concentration values from the two samples (first assay and last assay) and used the average over these 20 concentration values as our estimate for the actual cell concentration during all of the assays performed on that day. The standard error of the mean of these 20 concentrations was used as their error.

\subsection*{Transconjugant concentrations}
We used the time to threshold method~\cite{bethke2020environmental} to count transconjugants within the mix of donors, recipients and transconjugants obtained at the end of a conjugation assay. The method requires amplifying the population of transconjugants through selective growth and measuring the time the population takes to reach a prescribed OD threshold~(Fig.~\ref{fig:SI_GrowthCurves}B). This time-to-threshold is then converted into the starting cell concentration using a calibration curve~(Fig.~\ref{fig:SI_cellcount}F). We typically allocated nine wells (Fig.~\ref{fig:SI_GrowthCurves}A) on the plate to each conjugation assay. We performed the conversion from the measured times-to-threshold to transconjugant concentrations individually for each well, and only then computed the mean and standard error of the mean over the concentrations of these technical replicates.

To create the calibration curve (Fig.~\ref{fig:SI_cellcount}F), we isolated and cultured transconjugant cells. From this culture, a dilution series was created and for the individual dilutes, two parameters were measured: First, the cell concentration was determined using a hemocytometer. For the higher dilutions (cell concentration of $\SI{1.2e5}{\per\milli\liter}$ and lower), where the cell concentration got too small to be measured in the counting chamber, the value of the cell concentration was extrapolated from the other data points at lower dilutions (cell concentration of $\SI{4.2e5}{\per\milli\liter}$ and higher). For the extrapolation, the logarithm of the cell concentrations was plotted against the logarithm of the dilution, and a regression curve fitted onto that data. By extrapolating along this curve, the additional data points were then determined. Second, we loaded the dilutes each into 8 wells of a 96-well-plate containing the double selective LB medium and let them grow at $\SI{37}{\degree C}$ while the OD600 was measured every 5 min. From the measured growth curves, we determined the time it took the sample to pass the threshold optical density of 0.200. The best-fit trend line (linear regression between the logarithm of the cell concentrations and the times-to-threshold) displayed in Fig.~\ref{fig:SI_cellcount}F served as the calibration curve in all our experiments.

\section{Mathematical analysis of stochasticity in the time to threshold method}
The time to threshold method can be heuristically understood by assuming that the large-time population size is of the form $N(t)=N(0)\exp(\alpha t)$, where $\alpha$ is the population growth rate and $N(0)$ is the initial population size. Inverting this relationship enables one to estimate $N(0)$ from $N(t)$, where $N(t)$ is experimentally accessible. However, as can be seen from the calibration curve~(Fig.~\ref{fig:SI_cellcount}F) and typical growth curves in our experiments (Fig.~\ref{fig:SI_GrowthCurves}B), the times-to-threshold are increasingly more variable as the initial concentration of transconjugants decreases. This variability is, in part, generated by stochastic variability in cell division, raising the question of the limits of applicability of the time to threshold method at low starting concentrations. In this Section, we analyze the time to threshold method using the general theory of stochastic growth processes and we show that the method remains applicable at the starting cell concentrations of interest. We make only a mild set of mathematical assumptions: we assume that doubling times of cells are identically and continuously distributed across cells and that each cell behaves independently of other cells (including those in its lineage). The doubling times must have a finite mean, but no other assumptions are made on their distribution.

\subsection{Stochastic growth process}
We denote by $c(t)$ the concentration of (transconjugant) cells undergoing the growth process at time $t$, starting from some initial condition $c(0)$, and by $N(t)$ (resp. $N(0)$) the cell count. By the general theory of Bellman-Harris processes \cite{harris2002theory}, it follows that for large $t$,
\begin{equation}
N(t)\sim \beta e^{\alpha t}.
\label{eq-exponential}
\end{equation}
Here:
\begin{itemize}
\item $\alpha$ is a \emph{deterministic} growth rate, which is solely determined by the distribution of the doubling time and (provided no other sources of uncontrolled variability are present) \emph{does not vary between experiments} performed under identical conditions. Concretely, letting $f_D$ be the probability density governing the cell doubling time, $\alpha$ is the unique solution to the fixed point equation:
\begin{equation}
2\int_0^\infty \exp(-\alpha t) f_D(t)dt=1.
\end{equation}
For example, it is easy to check that when $f_D(x)=\lambda \exp(-\lambda x )$ (i.e., doubling times are exponentially distributed with rate $\lambda$), we obtain $\alpha=\lambda.$ More generally, when $f_D(x)=\lambda^{k} x^{k-1} \exp(-\lambda x )/\Gamma(k)$ (i.e., Gamma distribution with shape $k$ and rate $\lambda$) and $k$ is a positive integer, we have
\begin{equation}
\alpha=\lambda(2^{1/k}-1).
\end{equation}
\item $\beta$ is a \emph{random variable} whose distribution is determined by that of the doubling time and mean proportional to the number of cells present at the outset. Precisely, starting from a fixed number of cells, $N(0)$,
\begin{equation}
\mathbb E(\beta)=m_0N(0)\quad\quad\text{where}\quad\quad m_0=\frac{1}{4\alpha\int_0^\infty t \exp(-\alpha t) f_D(t) dt}.
\label{eq-beta}
\end{equation}
E.g. when $f_D(x)=\lambda \exp(-\lambda x)$, we obtain $m_0=1.$ When $f_D(x)=\lambda^k x^{k-1} \exp(-\lambda x)/\Gamma(k)$ where $k$ is a positive integer, 
\begin{equation}
m_0=\frac{(\alpha+\lambda)^{k+1}}{4k\alpha\lambda^k}.
\end{equation}
\end{itemize}
The convergence in \eqref{eq-exponential} is interpreted as convergence of the random variable $N(t)e^{-\alpha t}$ to the random variable $\beta$ both in mean square and (provided a mild technical condition) with probability 1, i.e., for every sample path. (Either condition will imply convergence in distribution.)

The role of the two parameters in the large-time exponential growth is better understood by taking the natural logarithm of the observable:
\begin{equation}
\log N(t)\sim \log \beta +  \alpha t,
\label{eq-exponential-log}
\end{equation}
showing that the growth rate $\alpha$ is the deterministic (unchanging) slope of the curve and $\log\beta$ is the offset, which is related to the delay in reaching the exponential growth. 

Figs.~\ref{fig:stochastic_analysis}A--C present a worked example of the full evolution of a stochastic growth process and the estimation of the various quantities. The example assumes the cell division times to be exponentially distributed with rate parameter 1 (division per time unit), although we stress that the general theory is not contingent on any particular choice of the doubling time distribution.
Within, the role of the deterministic growth rate $\alpha$ and the random offset $\beta$ are illustrated in Fig.~\ref{fig:stochastic_analysis}A, showing the simulated paths of the process assuming an exponential lifetime distribution. When further information can be extracted about the doubling times, the growth process can be characterized further. In the worked example of Fig.~\ref{fig:stochastic_analysis}B, the random offset $\beta$ is distributed as a Gamma random variable with shape parameter $N(0)$ and rate 1.

\subsection*{Calibration Curve}
We now consider the stochastic aspects of the calibration curve, explain how to use it to estimate $N(0)$ in the stochastic regime, and discuss its limits.

Let $\tau_1,\ldots,\tau_k$ be the times at which $k$ replicates of the growth process (with corresponding random offsets $\beta_1,\ldots,\beta_k$ and deterministic growth rate $\alpha$) reach some pre-determined threshold population size $\nu$. By Eq.~(\ref{eq-exponential-log}), we have
\begin{equation}
\log\nu=\log\beta_i+\alpha \tau_i
\label{eq-calibration-curve}
\end{equation}
for each measurement $i=1,\ldots,k$. In particular, irrespective of the choice of the threshold $\nu$,
\begin{equation}
\log\beta_i-\log\beta_j=-\alpha (\tau_i-\tau_j).
\end{equation}
See Fig.~\ref{fig:stochastic_analysis}C for an illustration. The mean of $\beta$ can now be estimated using the maximum likelihood (optimal) estimator as \begin{equation}\overline{\beta}:=\frac{\beta_1+\ldots+\beta_k}{k}.\label{eq-estimator-beta}\end{equation}
For example, we previously remarked that when the doubling times are exponentially distributed, $\beta$ is a Gamma random variable with shape parameter $N(0)$ and rate 1. In this scenario, we also have $m_0=1$, and the problem of estimating $N(0)$ is, therefore, equivalent to estimating the shape parameter of a Gamma distribution with a known rate parameter. Since the mean of a Gamma random variable is the product of the two parameters (the unknown shape and the fixed known rate), the solution is indeed as given in Eq.~(\ref{eq-estimator-beta}). 

More generally, \eqref{eq-calibration-curve} is at the heart of the time to threshold method we employ and the underlying `calibration curve', with one key modification. While measurements of $\beta$ are easily accessible in simulation, e.g. by estimating $\alpha$ from the sample paths of the process and letting $\beta=N(t) \exp(-\alpha t)$ for any fixed large time $t$, this is a more challenging problem in a microbial experiment. As the cell counts at a given time $t\neq 0$ are often measured by proxy (optical density in our experiments), the method is adapted to:
\begin{enumerate} \item[(1)] Work directly with the starting concentrations $N(0)$ rather than $\beta$.
\item[(2)] In the calibration phase, infer the parameters of the curve from calibration measurements based on known (fixed) values of $N(0)$.  
\item[(3)] In the experiment phase, use the same parameters to infer the values $N(0)$ from those of the times-to-threshold.
\end{enumerate}

To justify the shift from observing $\beta$ to observing $N(0)$, note that when $k$ is large, by Eq.~(\ref{eq-beta}), we have, 
\begin{equation}\overline{\beta}=\frac{\beta_1+\ldots+\beta_k}{k} \xrightarrow[k\to\infty]{\text{LLN}} N(0)m_0.
\end{equation}
Averaging \eqref{eq-calibration-curve},
\begin{equation}
\log\nu=\overline{\log\beta_i}+\alpha \overline{\tau_i}.
\label{eq-calibration-curve-2}
\end{equation}
By Jensen's inequality, 
\begin{equation}
\overline{\log(\beta)}\leq \log(\overline{\beta}).\label{eq-Jensen}\end{equation} 
However, when the distribution of $\beta$ is concentrated enough around its mean so that $\overline{\log(\beta)}\approx \log(\overline{\beta})$, we have available a different calibration curve:
\begin{equation}
\log\nu\approx\log\overline{\beta_i}+\alpha \overline{\tau_i}.
\end{equation}
For $k$ large, we expect to see
\begin{equation}
\log\nu-\log m_0\approx \log N(0)+\alpha \overline{\tau_i}.
\end{equation}
For example, we have remarked that when the cell doubling times are exponentially distributed and starting from a fixed cell count $N(0)$, $\beta$ is distributed as a Gamma$(N(0),1)$ random variable. In this case, Figure~\ref{fig:stochastic_analysis}D shows that $\log(\overline{\beta_i})$ and $\overline{\log \beta_i}$ are numerically very close for $N(0)=100$. Generally, both the mean and the variance of $\beta$ are proportional to $N(0)$. (This is because every cell initiates an independent growth process.) It follows that the bulk of the data is in a relatively concentrated band around the mean. Therefore, for $N(0)$ sufficiently large, as is the case at present, we expect $\log(\overline{\beta_i})$ and $\overline{\log \beta_i}$ to be relatively close. From the experimental perspective, given that a single well in a 96-well-plate holds a sample volume of about $V_\tn{well}=\SI{200}{\micro\liter}$, these calculations show that initial concentrations as low as $c(0)= N(0)/V_\tn{well}\approx \SI{5e2}{\per\milli\liter}$ can be accurately detected using the calibration curve.

We can now mathematically describe the calibration part of the experiment. Consider $\ell$ replicates of the process, grown from starting concentrations $N^{(1)}(0)=n^{(1)},\ldots, N^{(\ell)}(0)=n^{(\ell)}$  which are known and fixed, or potentially Poisson-distributed around known fixed means $n^{(1)},\ldots,n^{(\ell)}$. Furthermore, the starting values $n^{(1)},\ldots,n^{(\ell)}$ are selected to span the experimentally relevant range. For each $n^{(j)}$, we observe $k=8$ replicates of the process starting from $n^{(j)}$ (or from $n_1^{(j)},\ldots,n_k^{(j)}$ to allow for stochasticity in the starting values). We denote by $\tau_1^{(j)},\ldots,\tau_k^{(j)}$ the resulting times-to-threshold and use linear regression to obtain best fit of the form
\begin{equation}
y=\hat cx+\hat d,
\label{eq-regression}
\end{equation}
based on the data (cf.~Fig.~\ref{fig:SI_cellcount}F):
$$\left( \overline{\tau^{(1)}},\overline{n^{(1)}}\right),\ldots, \left(\overline{\tau^{(\ell)}},\overline{n^{(\ell)}}\right).$$
Subsequently, when $N(0)$ is the outcome of the encounter phase of the experiment and is no longer known, its mean is estimated from Eq.~(\ref{eq-regression}) with the parameters as estimated in the calibration phase. An example outcome of such a calibration phase for very low starting cell counts, where we expect most discrepancy with the theory, is shown in Fig.~\ref{fig:stochastic_analysis}E (where we plot the logarithm of $\beta/N(0)$ vs threshold times) and Fig.~\ref{fig:stochastic_analysis}F (plotting the logarithm of $N(0)$ vs threshold times). The plot suggests that the linear calibration curve might break down at initial concentrations on the order of $c(0)\approx \SI{10}{\per\milli\liter}$, and this low concentration regime requires further analysis.

\subsection*{Accounting for growth lag-time}

A further source of stochasticity in the growth process is due to the initial lag-time in cell growth. Mathematically, this is reflected in the time until the first division being differently distributed than the remaining doubling times. It is easy to show that the lag-time will not affect the growth rate $\alpha$ of the process. Indeed, the lag affects the division of only $N(0)$ cells. At $t$ sufficiently large, all of these cells will have divided. The process can then enter the `normal' growth phase, albeit with a delay. Figs.~\ref{fig:stochastic_analysis}G--J illustrate the impact of the lag phase on the calibration curve for two processes with the same doubling time distribution, one without initial lag~(Figs.~\ref{fig:stochastic_analysis}G--H) and one with initial lag~(Figs.~\ref{fig:stochastic_analysis}I--J). The exponential growth of the process will not be affected, but the delay will be reflected in the value of $\beta$, i.e., the population will grow as some $\beta'\exp(\alpha t)$. Since, as we have shown, the calibration curve approach estimates $\beta'$ directly from the data, the method remains valid and no additional work is required to compensate for the lag-time.

\section{Recovery from highest shear}
As shown in Figure 1D of the main text, the conjugation rate, just like the probability of having a successful conjugation at an encounter (Fig.~2C), decreases drastically at the highest shear. This begs the question about what is causing this sharp decrease. One possible hypothesis, which we reject in this section, is that the high shear forces damage the pilus or other parts of the T4SS, suppressing the total number of conjugation events that can occur.

    To test this hypothesis, we first subjected a mixture of donor (containing the R1-19 plasmid) and recipient cells (both prepared at OD600 = 0.05) to a shear rate of $\SI{1e3}{\per\second}$ for $\SI{30}{\minute}$ during a conjugation assay in the rheometer~(Fig~\ref{fig:SI_100_1000_100}A). After this step, a $\SI{200}{\micro\liter}$ sample was taken to determine its transconjugant concentration using the time to threshold method under double selection. However, in contrast to the usual conjugation assays, the mixture of donors and recipients in the rheometer was not discarded after this first assay, but the same mixture was reused immediately afterwards for a second conjugation assay at a shear rate of $\SI{1e2}{\per\second}$~(Fig~\ref{fig:SI_100_1000_100}A). After an additional 30 min of mixing at the lower shear rate, we collected another $\SI{200}{\micro\liter}$ sample to determine its transconjugant concentration with the time to threshold method. To have a point of comparison for this second assay, a third conjugation assay was performed at a shear rate of $\SI{1e2}{\per\second}$, this time with a fresh mixture of donors and recipients. This series of three assays was sequentially repeated three times with the same cultures.
    As before, we measured the initial concentrations of both donors and recipients in the hemocytometer. Additionally, the hemocytometer was also used to measure the cell concentration in the mixture of donors and recipients right before and right after each assay during one of the three series to assess growth during these assays. With cells now spending up to one hour in the rheometer, growth may start to have a non-negligible impact on the conjugation rate. To account for growth, instead of Eq.~(1) of the main text, we used the endpoint formula determined by Simonsen et al.~\cite{Simonsen1990}:
    \be
        \label{eq:eta_Simonsen}
        \eta = \Psi \ln{\left(1 + \frac{c_\tn{T}(t_\tn{m}) c_\tn{N}(t_\tn{m})}{c_\tn{R}(t_\tn{m}) c_\tn{D}(t_\tn{m})}\right)} \frac{1}{c_\tn{N}(t_\tn{m}) - c_\tn{N}(0)},
    \ee
    where $\eta$ is the conjugation rate, $\Psi$ is the growth rate (in this formula assumed to be the same for donors, recipients and transconjugants), $c_\tn{D}(t_\tn{m})$, $c_\tn{R}(t_\tn{m})$, and $c_\tn{T}(t_\tn{m})$ are the concentrations of donors, recipients and transconjugants, respectively, after a mating duration of $t_\tn{m}$, $c_\tn{N}(t_\tn{m})=c_\tn{D}(t_\tn{m})+c_\tn{R}(t_\tn{m})+c_\tn{T}(t_\tn{m})$ is the total cell concentration after a time of $t_m$, and $c_N(0)$ is the initial total cell concentration.
    The final concentrations of donors and recipients were projected from the initial concentrations using an exponential growth model:
    \be
        \label{eq:DR_growth}
        c_i(t_\tn{m}) = c_i(0) \exp{\left( \Psi  t_m \right)} ,~~~ i \in \{ \tn{D, R} \}.
    \ee
    The final donor and recipient concentrations of the first assay were directly used as the initial concentrations of the second assay. Similarly, the final transconjugant concentration from the first assay obtained by the time to threshold method was transferred to be the initial transconjugant concentration of the second assay, this way getting itself included into the value of $c_N(0)$ there:
    \be
        \label{eq:conc_transfer}
        c_{i}^\tn{second assay}(0) = c_{i}^\tn{first assay}(t_\tn{m}) ,~~~ i \in \{ \tn{D, R, T} \}.
    \ee
    The growth rate for this model was based on the results obtained from cell counting the mixture of donors and recipients (and transconjugants) right before and right after the conjugation assay and applying an exponential growth model:
    \be
        \label{eq:growth_rate}
        \Psi = \frac{1}{t_\tn{m}} \ln{\left( \frac{c_\tn{N}(t_\tn{m})}{c_\tn{N}(0)}\right)}.
    \ee
    This was done once for each of the three conjugation assays during the second run of this experiment. Both assays at shear rate $\SI{1e2}{\per\second}$ yielded similar growth rates [$\SI[separate-uncertainty = true]{2.16(33)e-2}{\per\minute}$ and $\SI[separate-uncertainty = true]{1.78(37)e-2}{\per\minute}$; mean $\pm$ propagated SEM], in contrast to the assay at shear rate $\SI{1e3}{\per\second}$ [$\SI[separate-uncertainty = true]{1.10(42)e-2}{\per\minute}$; mean $\pm$ propagated SEM]. Hence, the average of the growth rate of the two assays at shear rate $\SI{1e2}{\per\second}$ was used as value for $\Psi$ in the evaluation of all assays at that shear rate, as was the growth rate measured at $\SI{1e3}{\per\second}$ for the evaluation of all assays performed at that shear rate.

    Finally, we normalized the conjugation rates for each of the assays by the conjugation rate observed in the respective control assay with the fresh donor recipient mixture at shear rate $\SI{1e2}{\per\second}$. These results are presented in Fig.~\ref{fig:SI_100_1000_100}B, which shows that the conjugation rate fully recovers to the same level as in the control when being exposed to the weaker shear rate of $\SI{1e2}{\per\second}$ after the initial exposure to the highest shear. This indicates that, if any damage to the conjugation machinery occurred during the donor cell's exposure to the $\SI{1e3}{\per\second}$ shear rate, this damage is only short-lived. More likely, other mechanisms, such as short encounter duration or high shear forces, suppress the conjugation in this extremely high shear regime, as discussed in the main text.

\section{Plasmid transfer vs. spontaneous mutations}
Here, we give more details on the computation reported in the main text where we compared the hypothetical plasmid transfer rate among Roseobacters with the spontaneous mutation rate. Assuming Roseobacters represent a fraction $\beta$ of all cells, we estimate the plasmid transfer rate as follows (Eq.~(10) in the main text)
\be
\label{eq:plasmid_transfer_roseo}
\kappa(z)=\tn{plasmid transfer rate} = \eta[\dot\gamma(z)]c_\tn{Roseo}^2(z),
\ee 
where $c_\tn{Roseo}(z)=\beta c(z)$ is the concentration of Roseobacters and $c(z)$ is the total cell concentration. We compute the spontaneous mutation rate $\phi$ within the Roseobacter subpopulation as
\be
\label{eq:smr_roseo}
\phi(z)=\tn{spontenous mutation rate} = \mu s c_\tn{Roseo}(z),
\ee 
where $\mu$ is the growth rate and $s$ is the mutation rate per genome per generation. The ratio of the two rates is
\be
\label{eq:rates_ratio}
\frac{\kappa(z)}{\phi(z)}=
\frac{\eta[\dot\gamma(z)]\beta c(z)}{\mu s}.
\ee
Fig.~\ref{fig:PT_vs_SMR} plots this ratio as a function of $c$ and $\beta$ with $\mu=\SI{1}{\per\day}$, $s=2.5\times 10^{-3}$. We set $\eta=\SI{1e-7}{\milli\liter\per\hour}$, which corresponds to the maximum in Fig.~1D at the shear rate $\dot\gamma= \SI{1e2}{\per\second}$ and mimics conditions in ocean breaking waves~($z=0$).

\section{Statistical analysis}
In this section, we analyze in detail the variability in the technical and biological replicates in the sequential up/down series and simultaneous shear vs. no shear conjugation experiments, which correspond to the data shown in Fig.~1D, Fig.~2A (same data as in Fig.~1D) and Fig.~\ref{fig:SI_Up_Down_ramp}. We also perform statistical tests on the key result that shear increases the conjugation rate in the two types of experiments.

\subsection*{Technical replicates} The variability in the conjugation rate determined in a single conjugation assay arises from the variability in the measured concentrations of donors, recipients and transconjugants~(Eq.~1). 
To quantify the variability in cell concentrations, we computed their mean $\bar{c}_\alpha$ and standard error of the mean $s_{\bar{c}_\alpha}$:
\bse
    \label{eq:SEM}
\be
 \bar{c}_\alpha&=&\f{1}{N_\alpha}\sum_{j=1}^{N_\alpha} c_{\alpha,j}, \\
    s_{\bar{c}_\alpha} &=& \sqrt{\frac{1}{N_\alpha \left( N_\alpha-1 \right)} \sum_{j=1}^{N_\alpha} \left( c_{\alpha,j} - \bar{c}_\alpha \right)^2 },
    \ee
\ese
where $N_\alpha$ is the number of samples and $c_{\alpha,j}$ are the values obtained from the individual concentration measurements for each cell type [$\alpha$ standing for either donors (D), recipients (R), or transconjugants (T)]. Typically, $N_\tn{D}=N_\tn{R}=20$, which corresponds to the number of randomly placed counting boxes in the hemocytometer (SI Section 2 and Fig.~\ref{fig:SI_cellcount}A), and $N_\tn{T}=9$, which corresponds to the number of wells on the well-plate used in the time to threshold method~(SI Section 2 and Fig.~\ref{fig:SI_GrowthCurves}) for this respective assay.
The mean values $\bar{c}_\alpha$ and their standard errors $s_{\bar{c}_\alpha}$ for the sequential up/down series experiments are presented in Table~\ref{tbl:myLboro}.

The variability in the measured cell concentrations propagates into the error of the conjugation rate $s_\eta$ of the individual conjugation assays (the vertical error bars in Fig.~\ref{fig:SI_Up_Down_ramp}A,B), because computing $\eta$ involves taking products and ratios of these quantities~(Eq.~1). We computed $\eta$ using the mean concentrations and propagated the error as follows 
\bse
    \label{eq:s_eta_abs}
\be
       \label{eq:s_eta_abs_A}
    \eta&=&\bar{c}_\tn{T}/(t_\tn{m}\bar{c}_\tn{D}\bar{c}_\tn{R}),
    \\
    \label{eq:s_eta_abs_B}
    s_\eta&=& \eta\sqrt{ \left( \frac{s_{\bar{c}_\tn{T}}}{\bar{c}_\tn{T}} \right)^2 + \left( \frac{s_{\bar{c}_\tn{D}}}{\bar{c}_\tn{D}} \right)^2 + \left( \frac{s_{\bar{c}_\tn{R}}}{\bar{c}_\tn{R}} \right)^2}.
\ee
\ese
Eq.~(\ref{eq:s_eta_abs_B}) assumes independence of the random variables and follows from standard error propagation laws for multiplication and division~\cite{JCGM2008}. The duration of the assay $t_\tn{m}$, also known as the mating duration, was assumed to have a negligible error and was thus ignored in the error calculation.
In summary, the points in Fig.~\ref{fig:SI_Up_Down_ramp}A,B show $\eta$, and the vertical error bars show $s_\eta$.

Fig.~\ref{fig:SI_Up_Down_ramp}D shows the enhancement factor in simultaneous shear ($\dot\gamma=\SI{100}{\per\second}$) vs. no shear ($\dot\gamma=\SI{0}{\per\second}$) experiments. We computed the enhancement factor $f$ (by taking the ratio between the different final concentrations of transconjugants) and its propagated error $s_f$  as follows
\bse
    \label{eq:f_s_f}
    \be
    \label{eq:f}
f&=&\bar{c}_\tn{T}(\dot\gamma =100) / \bar{c}_\tn{T}(\dot\gamma =0), \\
    \label{eq:s_f}
s_f &=& f\sqrt{ \left( \frac{s_{\bar{c}_\tn{T}(\dot\gamma =100)}}{\bar{c}_\tn{T}(\dot\gamma =100)} \right)^2 + \left( \frac{s_{\bar{c}_\tn{T}(\dot\gamma =0)}}{\bar{c}_\tn{T}(\dot\gamma =0)} \right)^2},
\ee
\ese
where $\bar{c}_\tn{T}(\dot\gamma)$ and $s_{\bar{c}_\tn{T}(\dot\gamma)}$ are the mean and standard error of the mean of the transconjugant concentration at shear rate $\dot\gamma$ computed across the technical replicates from the wells in the well-plate  (see Table~\ref{tbl:myLboro2}).

\subsection*{Biological replicates} Fig.~\ref{fig:SI_Up_Down_ramp}A,B shows that the variability between the technical replicates of each individual assay (calculated according to Eq.~\ref{eq:s_eta_abs_B} above, vertical error bars in Fig.~\ref{fig:SI_Up_Down_ramp}A,B) is small compared to the variability between biological replicates (points and individual curves in Fig.~\ref{fig:SI_Up_Down_ramp}A,B). For this reason, we decided not to propagate the errors from the technical replicates any further. As points and vertical error bars in Fig. 1D and Fig. 2A, we reported the mean $\bar{\eta}(\dot{\gamma})$ and standard error of the mean $s_{\bar{\eta}(\dot{\gamma})}$ of the conjugation rate over the different biological replicates at a given shear rate calculated as: 
\bse
    \label{eq:SEM_eta}
\be
    \bar{\eta}(\dot{\gamma})&=&N_\tn{br}^{-1}\sum_{j=1}^{N_\tn{br}}\eta_j(\dot{\gamma}), \\
    s_{\bar{\eta}(\dot{\gamma})} &=& \sqrt{\frac{1}{N_\tn{br} \left( N_\tn{br}-1 \right)} \sum_{j=1}^{N_\tn{br}} \left( \eta_j(\dot{\gamma}) - \bar{\eta}(\dot{\gamma}) \right)^2 },
\ee
\ese
where $\eta_j(\dot{\gamma})$ are the conjugation rates measured in the individual biological replicates~(with each value calculated according to Eq.~\ref{eq:s_eta_abs_A}), and $N_\tn{br}=4$ is the number of biological replicates. 

\textit{Monte Carlo simulations.} In the previous paragraph, we computed the standard error of the conjugation rate in Eq.~(\ref{eq:SEM_eta}) across biological replicates by ignoring the variability in technical replicates. That is, we took the conjugation rates $\eta_j(\dot{\gamma})$ measured in individual biological replicates to be constant and equal to the mean values calculated according to Eq.~(\ref{eq:s_eta_abs_A}). To check the impact of the variability of the technical replicates onto the standard error computed across the biological replicates, we performed Monte Carlo simulations. Specifically, we computed Eq.~(\ref{eq:SEM_eta}) $n=\SI{1e6}{}$ times, each time drawing $\eta_j(\dot{\gamma})$ from a normal distribution with its mean and standard deviation equal to the mean and standard error of the technical replicates (computed according to Eq.~\ref{eq:s_eta_abs_A} and Eq.~\ref{eq:s_eta_abs_B}). We performed such simulations for the R1-19 plasmid for shear rates $\dot\gamma=\SI{0.1}{\per\second}$ (the plateau) and $\dot\gamma=\SI{1e2}{\per\second}$ (the peak). In both cases, the mean standard error of the conjugation rate in Eq.~(\ref{eq:SEM_eta}) calculated over the $n=\SI{1e6}{}$ runs increased by less than 20\%, confirming that the variability in biological replicates is the dominant source of variation in our experiments.

\subsection*{Statistical tests} 
\begin{figure*}[t!]
    \centering
    \includegraphics[width=1.0\textwidth]
    {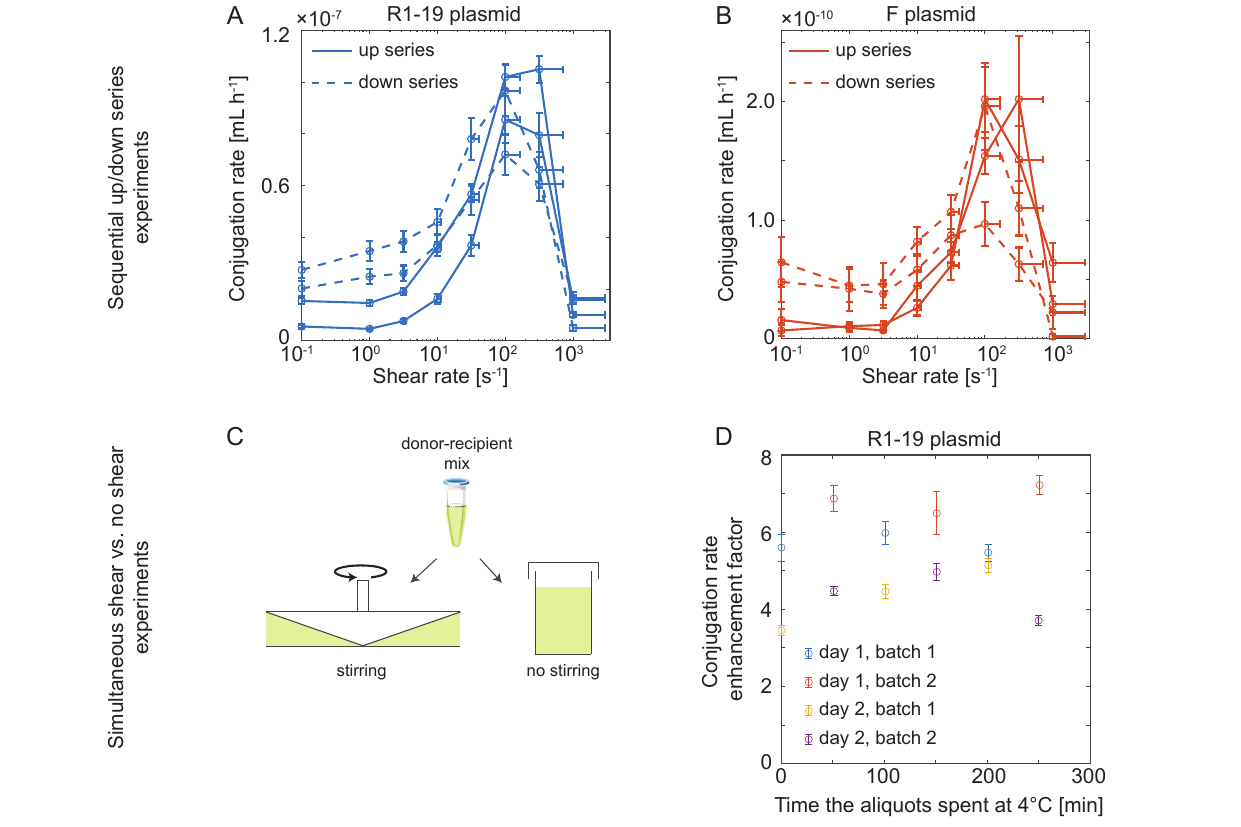}
    \caption{
    \textbf{Sequential up/down series and simultaneous shear vs. no shear conjugation experiments.} 
    \textbf{A, B,} 
    The curves in each graph depict independent conjugation experiments performed on different days (i.e., different biological replicates), four for the R1-19 plasmid (A) and four for the F plasmid (B). The lines are the same as the shaded lines in Fig. 1D. On a given day, we used the same cultures of donors and recipients and varied the shear rate between individual assays. The conjugation rate was measured from low to high shear rate twice (up series; solid lines) and from high to low shear rate twice (down series; broken lines). Points represent the averages~(Eq.~\ref{eq:s_eta_abs_A}) and vertical bars the propagated standard errors of the mean~(Eq.~\ref{eq:s_eta_abs_B}) from the technical replicates~(see SI Section 6). The one-sided horizontal error bars represent an additional shear rate above the baseline shear rate~($\dot\gamma_\tn{b}=\omega/\alpha$) generated by the secondary and turbulent flows at high rotation rates of the cone~(Eq.~\ref{eq:shear_additional}; see SI Section 1).
    \textbf{C,} Simultaneous shear vs. no shear experiments confirm the five-fold increase in the conjugation rate at the optimal shear for the R1-19 plasmid. We split a donor-recipient mixture into two samples of the same volume (1.2 mL). The first sample was exposed to shear $\dot\gamma=\SI{1e2}{\per\second}$ in the rheometer for 45 min at $\SI{30}{\degree C}$. The second sample was exposed to no shear by placing it in a stainless-steel container (same material as in the rheometer) in an incubator at $\SI{30}{\degree C}$. We then measured the concentration of transconjugants using the time to threshold method~(Table~\ref{tbl:myLboro2}), as in the sequential experiments.
    \textbf{D,} The enhancement factor computed as the ratio between the concentration of transconjugants in the sheared sample vs. the non-sheared sample. The experiment was performed on two days, using two different batches of donors and recipients on each day (a total of four biological replicates indicated in the figure through different colours). For each batch, three individual pairs of conjugation assays were performed, resulting in a total number of 12 pairs of conjugation assays. The individual assay pairs were performed sequentially, alternating between the two batches. The aliquots of donors and recipients that had initially been prepared and waited to be used were stored at $\SI{4}{\degree C}$. Hence, the aliquots for the first pair of assays did not spend any time at $\SI{4}{\degree C}$, whereas the aliquots for the last pair of assays were exposed to  $\SI{4}{\degree C}$ for about four hours.
    The time the aliquots spent at $\SI{4}{\degree C}$ (horizontal axis) did not affect the enhancement in the observed conjugation rate. Points represent the ratio of the average concentrations~(Eq.~\ref{eq:f}) and vertical bars the propagated errors~(Eq.~\ref{eq:s_f}) from $n=6$ technical replicates created from the sample of each assay, one for each well in a 96-well-plate.
    }
    \label{fig:SI_Up_Down_ramp}
\end{figure*}

For the sequential up/down series experiments, we tested the statistical significance of the impact of shear on the conjugation rate. Specifically, for each plasmid, we compared the $n_\tn{peak}=4$ values of the conjugation rate at shear rate $\dot\gamma=\SI{1e2}{\per\second}$ (the peak) with the $n_\tn{plateau}=4$ values of the conjugation rate at the lowest shear $\dot\gamma=\SI{0.1}{\per\second}$ (the plateau); these values correspond to the four biological replicates displayed in Fig.~1 and Fig.~\ref{fig:SI_Up_Down_ramp}A,B. We used the Kolmogorov-Smirnov one-sided two-sample test computed using the kstest2() function in MATLAB with the option ‘Tail’ set to ‘Larger’, which yielded, for both plasmids, the asymptotic $p$ value $p = 0.0055$ for the test statistic $k = 1$ (sample size $n_\tn{peak}=n_\tn{plateu}=4$). The rejection of the null hypothesis that the conjugation rates at the two shear rates come from the same distribution implies that the conjugation rate is greater at the optimal shear rate $\dot\gamma=\SI{1e2}{\per\second}$ compared to the shear rate $\dot\gamma=\SI{0.1}{\per\second}$.

Similarly, for the simultaneous shear vs. no shear experiments~(Figs.~\ref{fig:SI_Up_Down_ramp}C and D), we also tested the statistical significance of the impact of shear on the conjugation rate. We compared the tranconjugant concentrations measured at the shear rate $\dot\gamma=\SI{1e2}{\per\second}$ with the concentrations measured at shear rate $\dot\gamma=\SI{0}{\per\second}$. As samples, we used the $n=4$ biological replicates obtained by averaging the data shown in the rows in Table~\ref{tbl:myLboro2} over the three runs (i.e., each batch, averaged over the three runs, served as a single sample). We then used the Kolmogorov–Smirnov one-sided two-sample test computed using the kstest2() function in MATLAB with the option ‘Tail’ set to ‘Larger’, which yielded, for the R1-19 plasmid, the asymptotic $p$ value $p = 0.0055$ for the test statistic $k = 1$ (sample size $n_\tn{shear}=n_\tn{no shear}=4$).

\newpage

\newpage

\newpage

\newpage
\begin{figure*}[t!]
    \centering
    \includegraphics[width=1.0\textwidth]{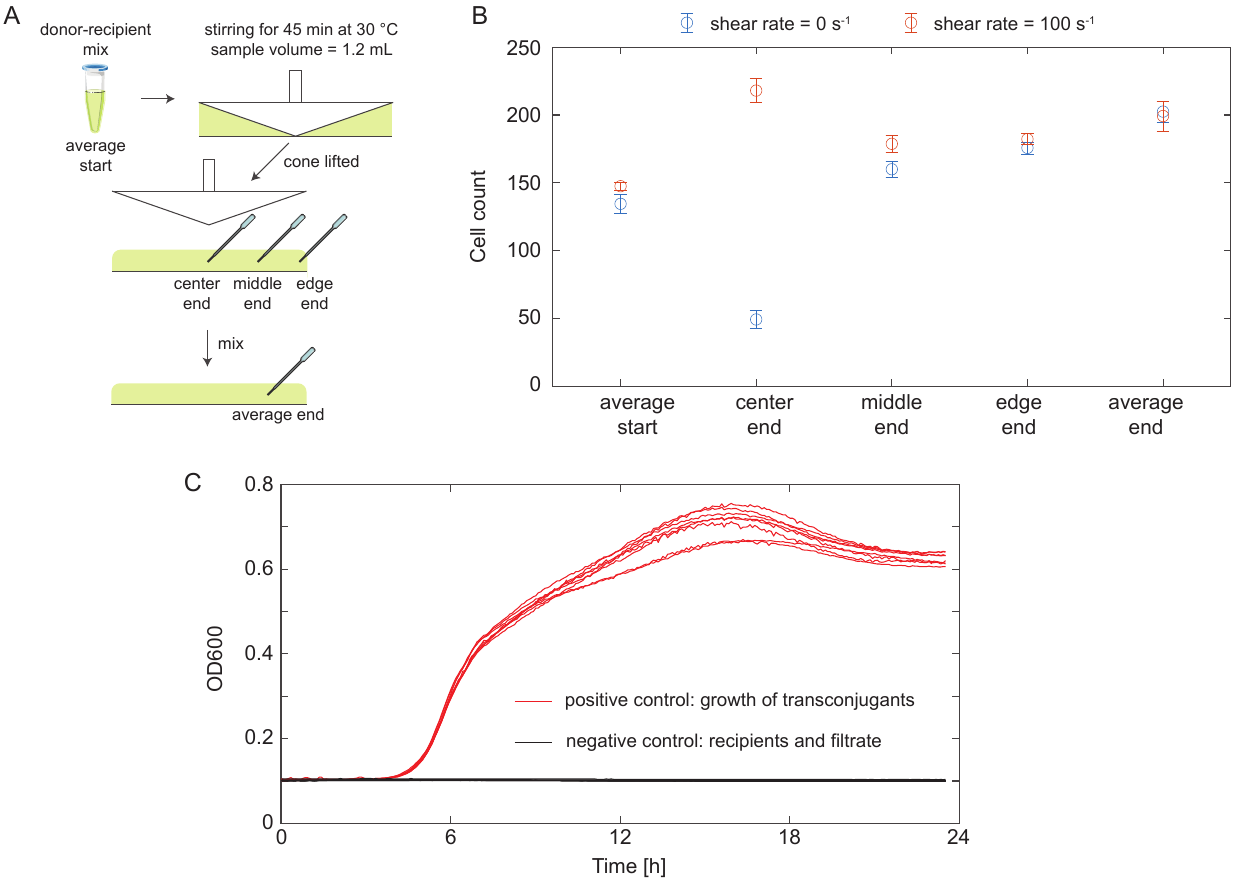}
    \caption{
    \textbf{Cells growth and distribution in the rheometer with and without shear and negative control for non-contact dependent plasmid transfer.}    
    \textbf{A,}
    To check the growth of the donor and recipient mix during the conjugation assay and the distribution of cells inside the rheometer at the end of the assay, we measured cell concentrations in sub-samples of that mixture. We investigated two assays: one exposed to the shear rate of $\SI{1e2}{\per\second}$ and the other to no shear for 45 minutes at $\SI{30}{\degree C}$. After stirring ended, we lifted the upper geometry and collected $\SI{50}{\micro\liter}$ from three positions: from the center, middle, and edge of the sample. Then, the remaining liquid in the rheometer was gently and thoroughly mixed by aspiration with the pipette and a fourth $\SI{50}{\micro\liter}$ sample was taken from this homogeneously distributed cell solution (average end). From all four of these samples, as well as from the initial donor-recipient mixture (average start), the cell concentration was determined using cell counting in a hemocytometer. 
    \textbf{B,} The data shows the average number of cells counted within one square of the hemocytometer (volume of $\SI{6.25}{\nano\liter}$; averaged over four individual squares, one per quadrant of the hemocytometer) for the samples taken at different positions inside the rheometer at the end of a conjugation assay, and the average samples from the start and the end. The error bars indicate the corresponding standard errors of the mean.
    The cell counts show that the cell concentration increases by about 40\% during the 30 min of the assay due to cell growth (average start vs. average end). Cells mixed at the shear rate of $\SI{1e2}{\per\second}$ are approximately uniformly distributed along the radial position with a slight accumulation at the cone center (10\% above the average). Cells exposed to no shear show a stronger depletion of cells near the center (75\% below the average), likely due to evaporation-driven accumulation~\cite{sempels2013auto,ruan2023evaporation}.
    \textbf{C,} We performed an additional control to eliminate the possibility that the plasmid transfer in our experiments occurs via non-contact horizontal gene transfer pathways, e.g., through the shedding of plasmids or extracellular vesicles by donors into the liquid, followed by their uptake by recipients from the liquid rather than through contact with donors. In this control, we first exposed a culture of donors (R1-19 plasmid; OD600=0.05; no recipients) to the optimal shear rate of $\SI{1e2}{\per\second}$ for 30 min at $\SI{30}{\degree C}$. We then filtered the suspension using a $\SI{0.45}{\micro\meter}$ filter and mixed the filtrate (containing hypothetical plasmids or vesicles but not donor cells, as confirmed under the microscope) with recipients (OD600=0.05), followed by exposure of the recipient-filtrate mix to the optimal shear rate of $\SI{1e2}{\per\second}$ for 30 min at $\SI{30}{\degree C}$. After that, we proceeded with the usual selective growth in the time to threshold method. The black lines in the graph show the corresponding OD600 readout for eight wells (technical replicates) on the plate, confirming that transconjugants did not form in the recipient-filtrate mixture. The red curves show the results of a positive control for a donor-recipient mixture (OD600=0.05 for both) after exposure to the optimal shear rate of $\SI{1e2}{\per\second}$ for 30 min at $\SI{30}{\degree C}$.
    }
    \label{fig:SI_cell_distribution}
\end{figure*}

\clearpage 
\newpage
\begin{figure*}[t!]
    \centering
    \includegraphics[width=1.0\textwidth]
    {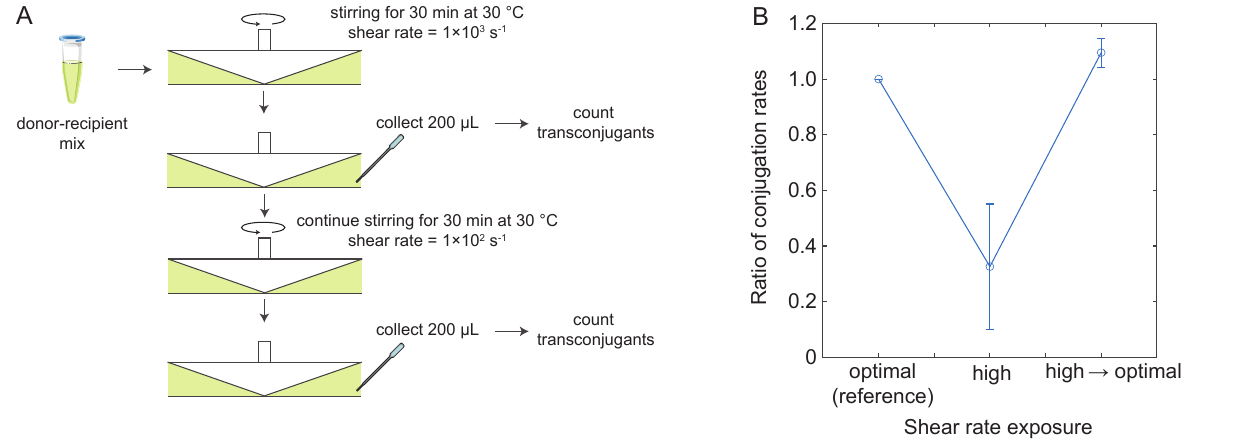}
    \caption{\textbf{Exposure to the high shear rate of $\SI{1e3}{\per\second}$ does not affect the long-term ability of donor cells to conjugate.} 
    \textbf{A,}
    To test that at the highest shear rate the shear did not irreversibly damage the cells or pili, we performed conjugation experiments with the R1-19 plasmid where we first exposed cells to the highest shear rate used in our experiments ($\dot\gamma=\SI{1e3}{\per\second}$) and then exposed them to the optimal shear rate ($\dot\gamma=\SI{1e2}{\per\second}$).     
    \textbf{B,}
    Ratios between the measured conjugation rates for the intervals described in (A) and a control (reference) experiment in which cells were exposed only to the optimal shear rate~($\dot\gamma=\SI{1e2}{\per\second}$). The points and error bars indicate the mean and standard error of the mean over $n=3$ biological replicates. These ratios show that cells fully recovered the same maximal conjugation rate at the optimal shear rate despite the initial exposure to the high shear rate. To account for growth, we used the endpoint formula to compute the conjugation rate~(Eq.~\ref{eq:eta_Simonsen} and SI Section 4).
    }
    \label{fig:SI_100_1000_100}
\end{figure*}

\clearpage 
\newpage
\begin{figure*}[t!]
    \centering
    \includegraphics[width=1.0\textwidth]
    {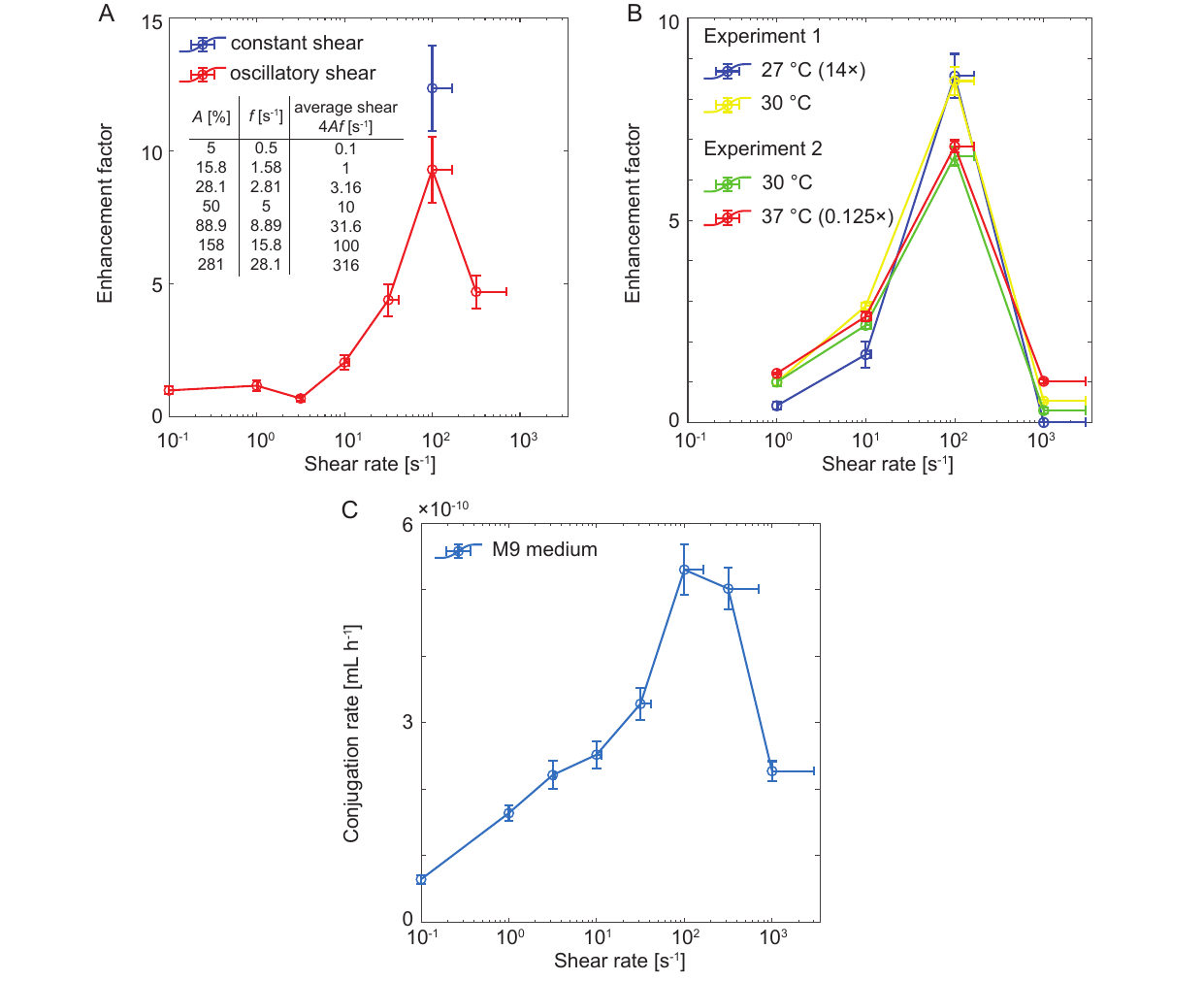}
    \caption{
    \textbf{The presence of a peak in the conjugation rate at optimal shear rate is robust to selective perturbations of experimental parameters.} 
    The figure shows the results of additional conjugation experiments with the R1-19 plasmid.
     \textbf{A,} Shear-induced enhancement in the conjugation rate as a function of shear rate for the oscillatory shear experiment. We performed a single conjugation experiment (up series; at $\SI{30}{\degree C}$) in oscillatory shear where the donor-recipient mix was exposed to the shear rate of the form $\dot\gamma(t)=2\pi A f\sin(2\pi f t)$, where $A$ is the strain amplitude and $f$ is the oscillation frequency. When averaged over a single oscillation period, the average shear rate experienced by the cells is $4Af$. These parameters were chosen to match, on average, the shear rate in the constant-shear experiments (inset Table). For reference, we show the value of the constant shear experiment at optimal shear performed with the same batch as the oscillatory shear experiment. All data points were normalized by the value of the conjugation rate at the lowest shear rate. Note that we could not increase the average shear rate further due to overheating of the rheometer.
     \textbf{B,} Shear-induced enhancement in the conjugation rate as a function of shear rate for two experiments performed at different temperatures. In each experiment, we carried out a pair of sequential up series experiments at four different shear rates and two different sets of temperatures ($\SI{27}{\degree C}$ or $\SI{37}{\degree C}$ together with a sweep at the reference temperature of $\SI{30}{\degree C}$; a total of eight conjugation assays per experiment). All data points were normalized by the value of the conjugation rate at the lowest shear rate for the corresponding reference temperature of $\SI{30}{\degree C}$. Changes in temperature only rescaled the response curve by a multiplicative prefactor shown in the legend without affecting the optimal shear rate value ($\dot\gamma=\SI{1e2}{\per\second}$).
     \textbf{C,} We performed additional conjugation experiments in M9 minimal medium (at $\SI{30}{\degree C}$) for cells grown in LB medium to mimic a sudden exposure to a low-nutrient environment. In M9 medium, the conjugation yield was about two orders of magnitude lower than in the LB medium, but the shear response curve retained the peak's position at the optimal shear rate.
     (A--C), All panels show a single replicate. Points and vertical error bars represent averages and standard errors of the mean over technical replicates (wells in the 96 well-plate).  As before, the one-sided horizontal error bars represent an additional shear rate above the baseline shear rate~($\dot\gamma_\tn{b}=\omega/\alpha$) generated by the secondary and turbulent flows at high rotation rates of the cone and assuming constant rotation rate~(Eq.~\ref{eq:shear_additional}; see SI Section 1).
    }
    \label{fig:SI_100_1000_100}
\end{figure*}

\clearpage
\newpage
\begin{figure*}[t!]
    \centering
    \includegraphics[width=1.0\textwidth]
    {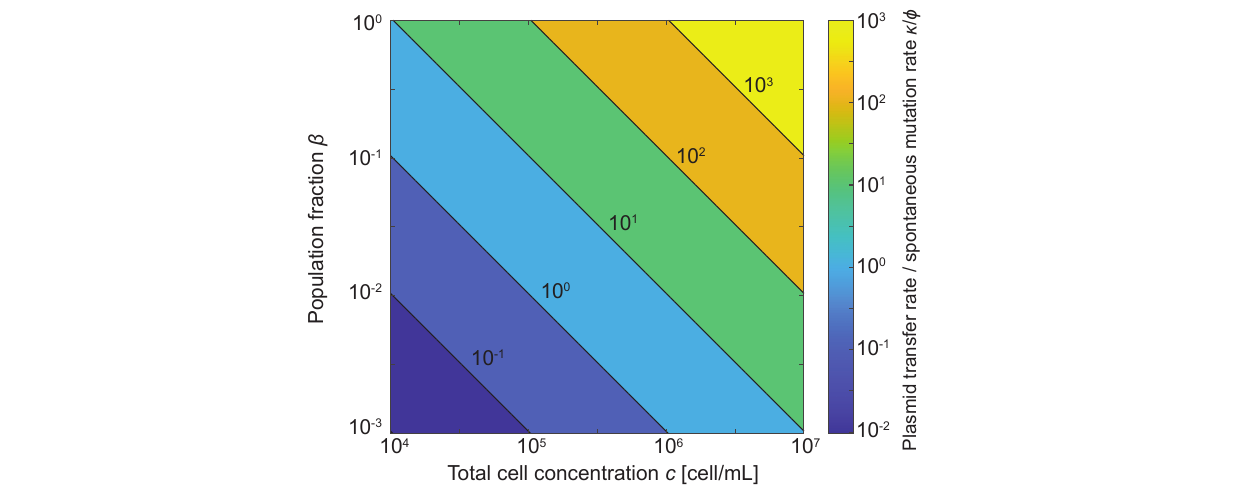}
    \caption{
    The ratio between the plasmid transfer rate and the spontaneous mutation rate calculated according to Eq.~(\ref{eq:rates_ratio}) as a function of the total cell concentration and fraction of the population participating in the plasmid transfer. Parameters used: growth rate $\mu=\SI{1}{\per\day}$, mutation rate per genome per generation $s=2.5\times 10^{-3}$ and conjugation rate $\eta=\SI{1e-7}{\milli\liter\per\hour}$, which corresponds to the maximum in Fig.~1D at the shear rate $\dot\gamma= \SI{1e2}{\per\second}$.
    }
    \label{fig:PT_vs_SMR}
\end{figure*}

\clearpage
\newpage
\begin{figure*}[t!]
    \centering
    \includegraphics[width=1.0\textwidth]
    {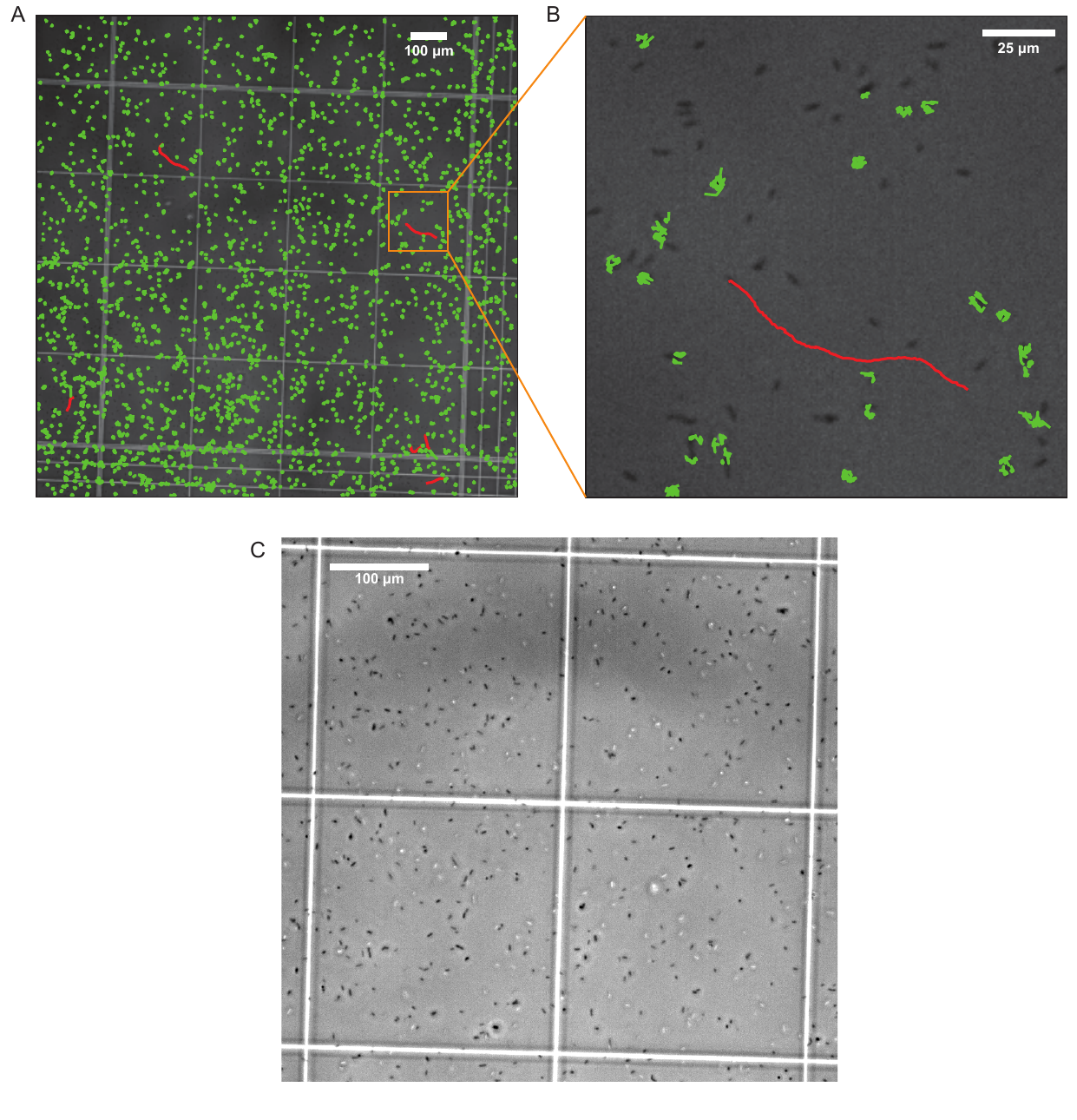}
    \caption{
    \textbf{Cell motility does not play a significant role in generating cell-cell encounters and cells do not substantially aggregate in our experiments.}
    \textbf{A,} Cell tracking in a donor-recipient mix shows that less than 1\% of cells were motile in our experiments, as expected, because the recipient strain lacks flagella, whereas donors lose flagella after centrifugation. We created a timelapse video of a donor-recipient mix imaged in the hemocytometer in the brightfield mode at 10x magnification for 10 s at 13 fps. We then tracked cells using Trackpy~\cite{allan_2024_12708864}. The tracks were classified as motile (red) if the track's end-to-end distance was at least $\SI{20}{\micro\meter}$ during the ten seconds; otherwise, the track was classified as non-motile and the observed movement was due to Brownian motion of suspended cells. Six tracks were classified as motile, whereas 2831 tracks as non-motile, which yields a motile fraction of 0.2\%.
    \textbf{B,} Zooming in onto a single motile cell surrounded by cells performing Brownian motion as well as cells stuck to the glass. Cells stuck to the glass were not tracked (dark spots), so the actual motile fraction is lower than 0.2\%.
    \textbf{C,} Microscopy image (brightfield, 10x magnification) of the donor-recipient mix in the hemocytometer. Shown is the sample taken from the rheometer after 30 min of stirring at the optimal shear rate ($\dot\gamma=\SI{1e2}{\per\second}$) for the R1-19 plasmid. No substantial aggregation of cells is observed.
    }
    \label{fig:SI_cell_motility}
\end{figure*}

\clearpage
\newpage
\begin{figure*}[t!]
    \centering
    \includegraphics[width=1.0\textwidth]
    {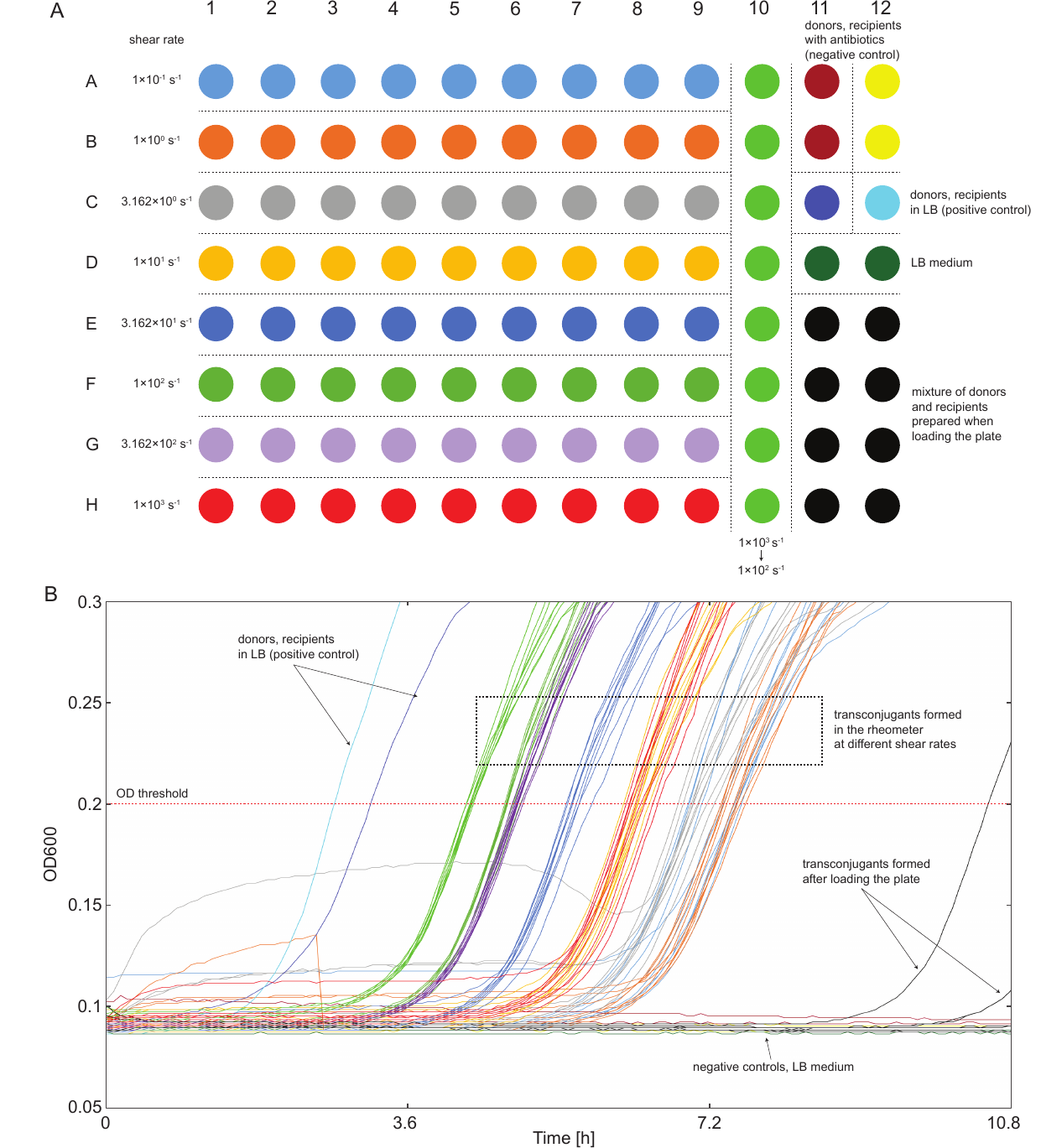}
    \caption{
    \textbf{Example of plate organization and growth curves in a single experiment.}
    \textbf{A,} Typical plate organization used. Technical replicates are shown in the same color. Alongside the samples coming from experiments corresponding to different shear rates, we also measured positive controls (donors, recipients in LB; empty LB medium) and negative controls (donors, recipients in antibiotics). In addition, eight wells were devoted to quantifying the formation of transconjugants in the plate reader (black); these wells were filled at the time of loading the plate with a freshly prepared donor-recipient mix with the same cell concentrations as in the experiment.
    \textbf{B,} The measured growth curves in a single (up series) experiment with the R1-19 plasmid. Positive controls confirm the viability of donors and recipients. Negative controls confirm that the antibiotics suppressed the growth of donors and recipients. LB medium control confirms no contamination occurred during the experiment. Transconjugants formed in the rheometer at different shear rates clearly separate into distinct bundles that cross the OD threshold of 0.2 at different times; these times-to-threshold were then converted to the concentrations of transconjugants using the calibration curve shown in Fig.~\ref{fig:SI_cellcount}F. Finally, the black curves show the growth of transconjugants that formed in the plate reader alone; they form very late, implying that their impact on the estimated concentrations of transconjugants formed in the rheometer is negligible.
    }
    \label{fig:SI_GrowthCurves}
\end{figure*}

\clearpage
\newpage
\begin{figure*}[t!]
    \centering
    \includegraphics[width=0.75\textwidth]
    {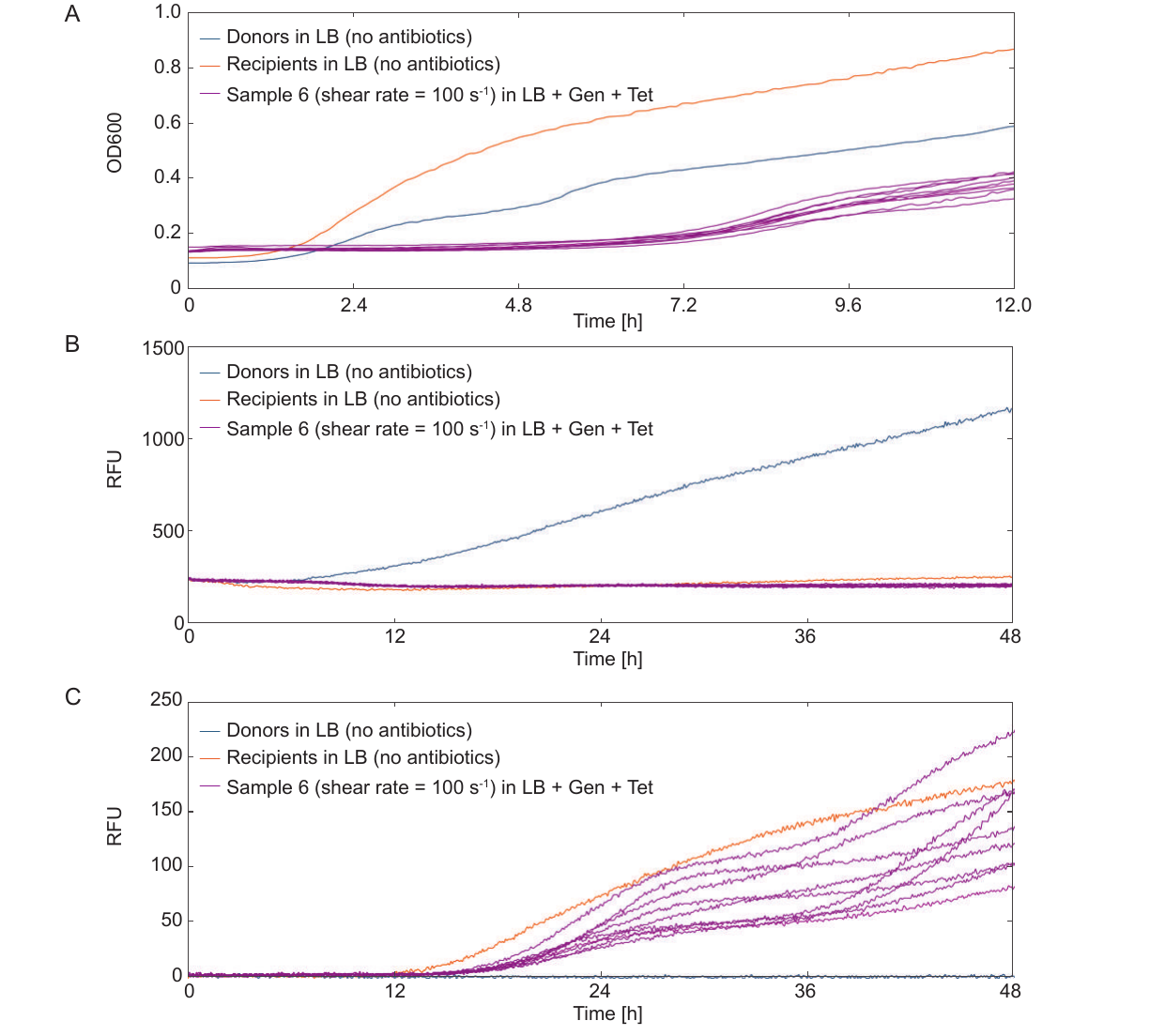}
    \caption{\textbf{Spectroscopic measurements provide additional verification of expected cell growth during the time to threshold method in the plate reader.} Independently from selection with antibiotics, chromosomal tagging of cells with fluorescent markers confirms that donors, recipients, and transconjugant cells grow as expected during the time to threshold method in the plate reader. In a selected experiment with the F plasmid, we additionally monitored the expression intensity of two fluorophores, dsRed (located on the chromosome of the recipients; Methods) and GFP (located on the chromosome of the donors; Methods).
    \textbf{A,} The OD600 as a function of time for donors and recipients grown in LB without antibiotics as positive controls for cell viability, and nine samples collected from the rheometer grown in a selective medium corresponding to the conjugation assay (for an F-plasmid) sheared at the shear rate of $\SI{1e2}{\per\second}$.
    \textbf{B,C,} The corresponding readouts in the green (B; excitation: 485/20 nm, emission: 528/20 nm) and red (C; excitation: 485/20 nm, emission: 590/35 nm) fluorescence channels are as expected: donors are present in the green but not the red channel, whereas recipients and transconjugants are present in the red but not the green channel. The expression of dsRed is only detectable in the stationary phase, as described in~\cite{choi2006mini}. Note the different range of values on the time axis (x-axis) in panel A vs. panels B and C.
    }
    
    \label{fig:SI_Fluorophores}
\end{figure*}

\clearpage
\newpage
\begin{figure*}[t!]
    \centering
    \includegraphics[width=1.0\textwidth]
    {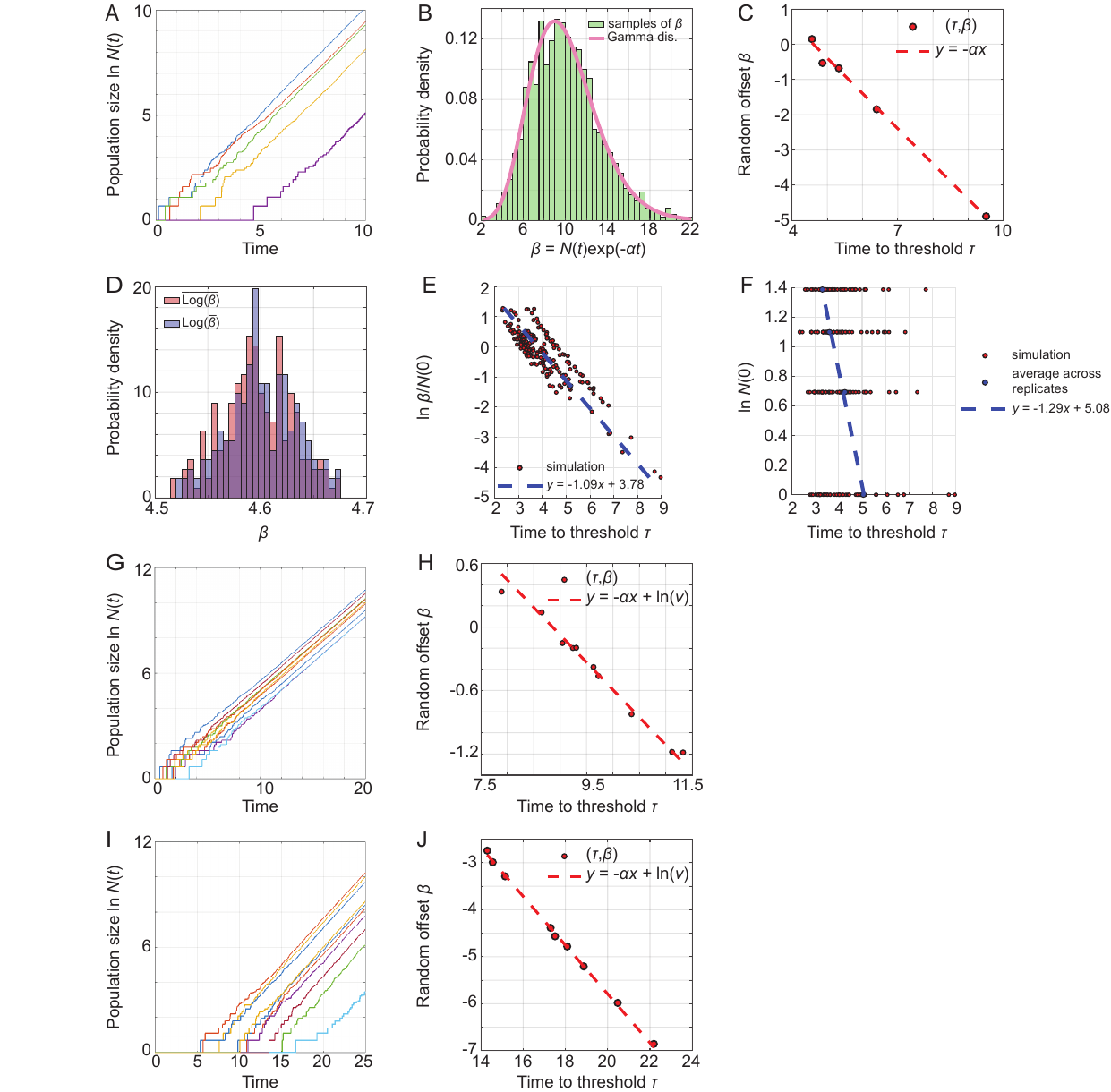}
    \caption{\textbf{Stochastic analysis of the time to threshold method.} SI Section 3 details the mathematical analysis of the impact of stochastic cell division on the time to threshold method.   
    \textbf{A-C}, Example of stochastic cell division where the cell doubling times are taken to be
    exponentially distributed with rate parameter 1.
    \textbf{A}, Sample paths of the growth process starting from $N(0)=1$ cell.
    \textbf{B}, Distribution of $\beta$ starting from $N(0)=10$ cells based on 2000 replicates and the theoretical prediction Gamma(10,1).
    \textbf{C}, $k=5$ measurements falling on the calibration curve (Eq.~\ref{eq-calibration-curve}), where the threshold was set to $\nu=100$.
    \textbf{D}, For exponentially distributed cell doubling times and the growth process starting from $N(0)=100$, $\beta$ is Gamma(100,1) distributed. The figure shows the difference between $\log(\overline{\beta_i})$ and $\overline{\log \beta_i}$. Each average is formed of 8 measurements, similar to the experimental setup. The histogram is formed of 2000 such averages.
    \textbf{E}, Simulation of the calibration experiment starting from small population values (leading to most uncertainty in outcomes). Cell doubling times are taken to be exponential with rate 1. The initial population values are deterministic and equal to $n_1=1,\ldots,n_4=4$. Each initial population size gives rise to $k=50$ replicates. The scatter plots the resulting values of $\log \beta$ against the times-to-threshold $\tau$. The slope is close to the theoretical slope $-\alpha=-1$.
    \textbf{F}, Same scenario as in panel (E), instead plotting $\log N(0)$ vs $\tau$, where $N(0)=1,2,3,4$. Jensen's inequality (Eq.~\ref{eq-Jensen}) is reflected in the slope deviating from the theoretical slope $\alpha=1$.
    \textbf{G, H}, Examples of the growth process and the calibration curve when the cell doubling times are Gamma distributed with shape parameter 3 and rate 2. The process starts from $N(0)=1$ cell, where $\nu=100$.
    \textbf{I, J}, Same scenario as in panels (G) and (H), except for the initial lag in division times. The lag time is Gamma
    distributed with shape parameter 10 and rate 1.
    }
    \label{fig:stochastic_analysis}
\end{figure*}

\clearpage
\newpage

  \begin{table}[h!]
    \begin{center}
      \begin{tabular}{  c }
        \includegraphics[width=1.0\textwidth]{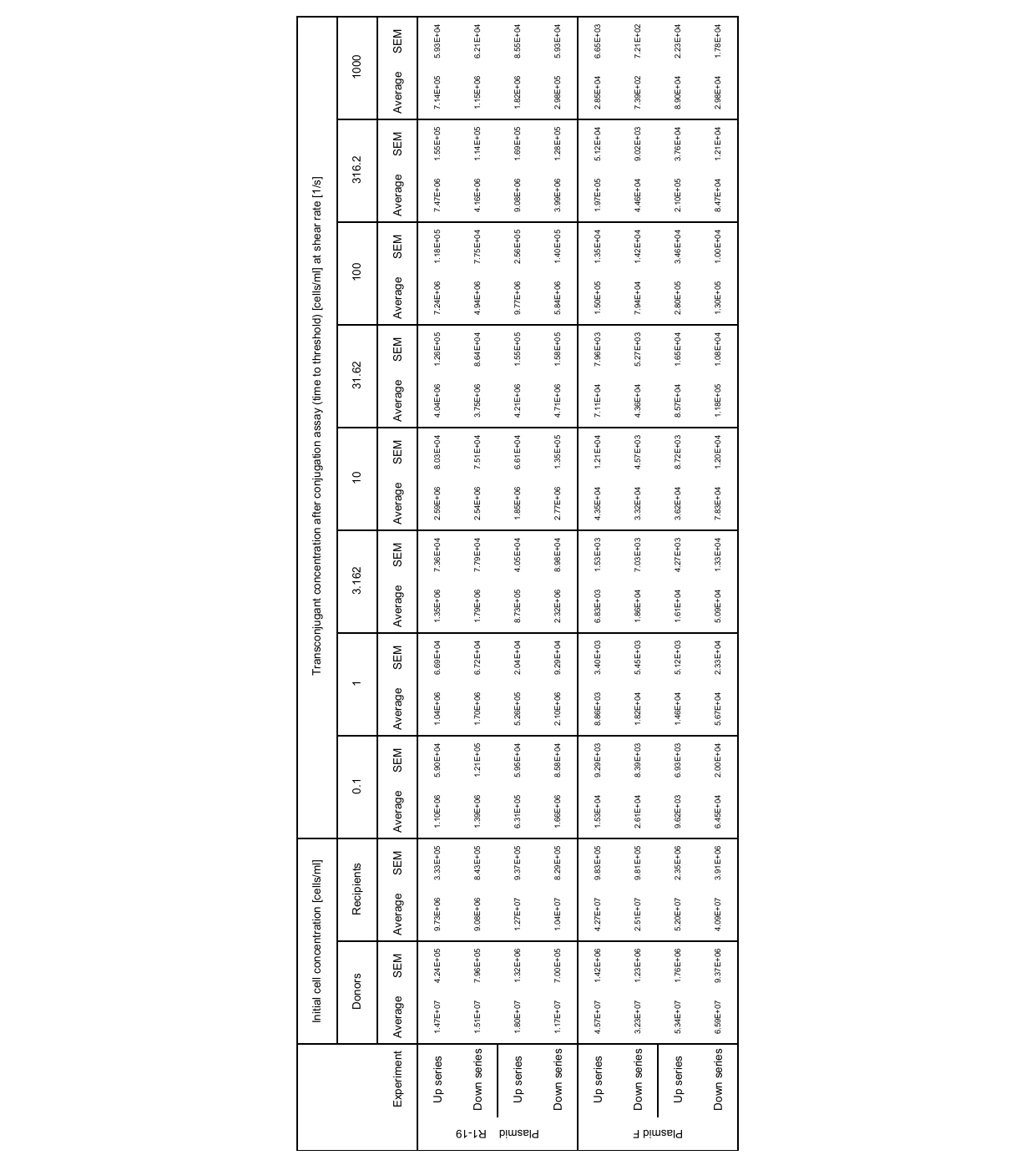}
      \end{tabular}
      \caption{
        Summary of measured concentrations of donors, recipients and transconjugants in the up/down series conjugation experiments.
        {\normalfont
        The rows correspond to experiments performed on different days (i.e., different biological replicates).
        The averages and standard errors of the mean (SEM) in each entry are computed over technical replicates corresponding to randomly placed counting boxes in the field of view of the hemocytometer (donors, recipients; typically $N_\tn{D}=N_\tn{R}=20$) and different wells in the well-plate (transconjugants; $N_\tn{T}=9$; SI Section 2).
        }
      }
      \label{tbl:myLboro}
    \end{center}
  \end{table}

  \begin{table}[h!]
    \begin{center}
      \begin{tabular}{  c }
        \includegraphics[width=1.0\textwidth]{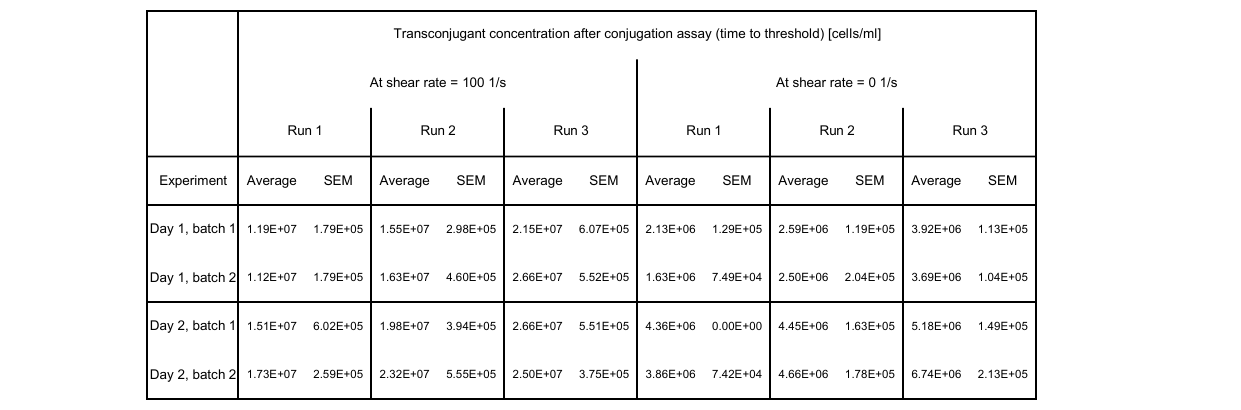}
      \end{tabular}
      \caption{
        Summary of measured concentrations of transconjugants in the simultaneous shear vs. no shear experiments for the R1-19 plamid.
        {\normalfont
        We performed two experiments on two days, each time with two different batches of both donors and recipients (rows; each batch represents a different biological replicate). Three runs were performed for each batch sequentially and alternating between the two batches. Shortly before each run, the donor-recipient mixture was prepared and split into two conjugation assays performed at the same time: one exposed to a shear rate of $\SI{1e2}{\per\second}$ and the other left still and unmixed~(Fig.~\ref{fig:SI_Up_Down_ramp}C).
        The data shown in the table represents the averages and standard errors of the mean (SEM) of the transconjugant concentrations obtained from these assays computed over the technical replicates corresponding to different wells in the well-plate ($N_\tn{T}=6$). We then used these values to compute the conjugation rate enhancement factor shown in Fig.~\ref{fig:SI_Up_Down_ramp}D by taking the ratio between the transconjugant concentrations with and without shear for each run. 
        In all experiments, donors and recipients were prepared at concentrations corresponding to OD600=0.05.
      }
      }
      \label{tbl:myLboro2}
    \end{center}
  \end{table}

\bibliographystyle{ieeetr}
\bibliography{biblio_Jonasz}

\end{document}